\documentclass[11pt,onecolumn,draftcls]{IEEEtran} %

\usepackage[linesnumbered, algoruled, longend, boxed]{algorithm2e}
\usepackage[dvips]{graphicx}
\usepackage{subfigure}
\usepackage{amsmath}
\usepackage{float}
\usepackage{amssymb}
\usepackage{bm}
\usepackage{amsthm}
\usepackage{color}
\usepackage{cite}

\newtheorem{theorem}{Theorem}

\newtheorem{proposition}{Proposition}
\newtheorem{corollary}{Corollary}

\newtheorem{lemma}{Lemma}

\newcommand{\stack}[3]{\mathrel{\mathop{\kern0pt #1}\limits_{#3}^{#2}}}

\def\x{{\mathbf x}}

\def\A{{\mathbf A}}
\def\B{{\mathbf B}}
\def\Z{{\mathbf Z}}
\def\T{{\mathbf T}}

\def\W{{\mathbf W}}
\def\I{{\mathbf I}}
\def\u{{\mathbf u}}

\def\J{{\mathbf J}}

\def\1{{\mathbf 1}}
\def\N{{\mathcal N}}

\def\1{{\bm 1}}

\newcommand{\bPhi}{\mathbf{\Phi}}

\newcommand{\bZ}{\mathbf{0}}
\newcommand{\bM}{\mathbf{M}}
\newcommand{\bC}{\mathbf{C}}
\newcommand{\bK}{\mathbf{K}}

\newcommand{\bv}{\mathbf{v}}
\newcommand{\bu}{\mathbf{u}}

\newcommand{\asym}{\textrm{asym}}
\newcommand{\con}{\textrm{c}}
\newcommand{\E}{\mathbb{E}}

\newcommand{\bg}{\mathbf{g}}

\newcommand{\bx}{\mathbf{x}}
\newcommand{\bA}{\mathbf{A}}

\newcommand{\bI}{\mathbf{I}}
\newcommand{\bJ}{\mathbf{J}}

\newcommand{\bW}{\mathbf{W}}
\newcommand{\bX}{\mathbf{X}}

\renewcommand{\Re}{\mathbb{R}}

\begin{document}

\title{Optimization and Analysis of Distributed Averaging with Short Node Memory}

\author{{Boris N. Oreshkin, Mark J. Coates and  Michael G. Rabbat} \\
Telecommunications and Signal Processing--Computer Networks
Laboratory\\
Department of Electrical and Computer Engineering \\
McGill University, Montreal, QC, Canada \\
Email: boris.oreshkin@mail.mcgill.ca, {\{mark.coates,
michael.rabbat\}@mcgill.ca} }

\maketitle
\vspace{-1cm}
\begin{abstract}

  \emph{Distributed averaging} describes a class of network algorithms
  for the decentralized computation of aggregate statistics.
  Initially, each node has a scalar data value, and the goal is to
  compute the average of these values at every node (the so-called
  \emph{average consensus} problem).  Nodes
  iteratively exchange information with their neighbors and perform
  local updates until the value at every node converges to the initial
  network average.  Much previous work has focused on algorithms where
  each node maintains and updates a single value; every time an update
  is performed, the previous value is forgotten.  Convergence to the average consensus is achieved
  asymptotically.  The convergence rate is fundamentally limited by
  network connectivity, and it can be prohibitively slow on topologies
  such as grids and random geometric graphs, even if the update rules
  are optimized.  In this paper, we provide the first theoretical
  demonstration that adding a local prediction component to the update
  rule can significantly improve the convergence rate of distributed
  averaging algorithms. We focus on the case where the local
  predictor is a linear combination of the node's current and previous values
  (i.e., two memory taps), and our update rule computes a combination
  of the predictor and the usual weighted linear combination of values
  received from neighbouring nodes.  We derive the optimal mixing
  parameter for combining the predictor with the neighbors' values,
  and conduct a theoretical analysis of the improvement in
  convergence rate that can be achieved using this acceleration
  methodology. For a chain topology on $N$
  nodes, this leads to a factor of $N$ improvement over standard
  consensus, and for a two-dimensional grid, our approach achieves a
  factor of $\sqrt{N}$ improvement.

\end{abstract}

\begin{keywords}

Distributed signal processing, average consensus, linear prediction.

\end{keywords}

%\clearpage
\section{Introduction}
\label{section::Introduction}

Distributed algorithms for solving the average consensus problem
have received considerable attention in the distributed signal
processing and control communities recently, due to their
applications in wireless sensor networks and distributed control of
multi-agent
systems\cite{
Xiao05,Moallemi06,Spanos05,Scherber04,Olfati-Saber04,Tang07,Ren05}.
See~\cite{Saber07} for a survey.  In the average consensus problem,
each node initially has a value, e.g., captured by a sensor, and the
goal is to calculate the average of these initial values at every
node in the network under the constraint that information can only
be exchanged locally, between nodes that communicate directly.

This paper examines the class of synchronous distributed averaging
algorithms that solve the average consensus problem. In this
framework, which can be traced back to the seminal work of
Tsitsiklis~\cite{Tsitsiklis84}, each node maintains a local estimate
of the network average. In the simplest form of a distributed
averaging algorithm, one iteration consists of having all nodes
exchange values with their neighbors and then update their local
average with a weighted linear sum of their previous estimate and
the estimates received from their neighbors.  This update can be
expressed as a simple recursion of the form $\bx(t+1) = \bW \bx(t)$,
where $x_i(t)$ is the estimate after $t$ iterations at node $i$, and
the matrix $\bW$ contains the weights used to perform updates at
each node.  (Note, $W_{i,j} \neq 0$ only if nodes $i$ and $j$
communicate directly, since information is only exchanged locally at
each iteration.)  Xiao and Boyd~\cite{Xiao04} prove that, so long as
the matrix $\bW$ satisfies mild contraction conditions, the values
$x_i(t)$ converge asymptotically to the initial average, as $t
\rightarrow \infty$.  However, Boyd et al.~\cite{Boyd06} have shown
that for important network topologies --- such as the two-dimensional grid or
random geometric graph, which are commonly used to model
connectivity in wireless networks --- this type of distributed
averaging can be prohibitively slow, even if the weight matrix
is optimized, requiring a number of
iterations that grows quickly with network size.

Numerical simulations have demonstrated that {\em predictive
  consensus} algorithms can converge much faster~\cite{Scherber04,Cao06,Ays09,Kok09,Joh08}. These algorithms
employ local node-memory, and change the algorithm so that the
state-update becomes a mixture of a network-averaging and a
prediction.  But there has been no
theoretical proof that they provide better performance, nor has
there been any analytical characterization of the improvement they
can provide. In addition, the algorithms have required intensive
initialization to calculate their parameters. In this paper, we
provide the first theoretical results quantifying the improvement
obtained by predictive consensus over standard memoryless consensus
algorithms. We focus on a linear predictor and derive a closed-form
expression for the optimal mixing parameter one should use to
combine the local prediction with the neighbourhood averaging. We
analytically characterize the convergence rate improvement and describe a simple decentralized
algorithm for initialization.

\subsection{Related Work}
\label{subsection::Related Work}

Two major approaches to accelerating the convergence of consensus
algorithms can be identified: optimizing the weight
matrix~\cite{Xiao04,Xiao05,Boyd06,Olfati-Saber04}, and incorporating
memory into the distributed averaging
algorithm~\cite{Ays09,Kok09,Joh08,Cao06,Scherber04,Sundaram07}. The
spectral radius of the weight matrix governs the asymptotic
convergence rate, so optimizing the weight matrix corresponds to
minimizing the spectral radius, subject to connectivity
constraints~\cite{Xiao04,Xiao05,Boyd06}. Xiao et al. formulate the
optimization as a semi-definite problem and describe a decentralized
algorithm using distributed orthogonal
iterations~\cite{Xiao04,Xiao05,Boyd06}. Although elegant and
efficient, this approach involves substantial initialization costs,
and the improvement does not scale in grid or random geometric graph
topologies (the averaging time is improved by a constant factor).

A more promising research direction is based on using local node
memory. The idea of using higher-order eigenvalue shaping filters was
discussed in~\cite{Scherber04}, but the problem of identifying
optimal filter parameters was not solved.  In~\cite{Cao06}
Cao \emph{et al.} proposed a memory-based acceleration framework for
gossip algorithms where updates are a weighted sum of previous state
values and gossip exchanges, but they provide no solutions or
directions for weight vector design or optimization.  Johansson and
Johansson~\cite{Joh08} advocate a similar scheme for distributed
consensus averaging.  They investigate convergence conditions and use
standard solvers to find a numerical solution for the optimal weight
vector. Recently, polynomial filtering was introduced for consensus
acceleration, with the optimal weight vector again determined
numerically~\cite{Kok09}. Analytical solutions for the
topology-dependent optimal weights have not been considered in
previous work~\cite{Kok09,Joh08,Cao06,Ays09} and, consequently, there
has been no theoretical convergence rate analysis for variants of
distributed averaging that use memory to improve the convergence rate.

Aysal et al. proposed the mixing of neighbourhood averaging with a
local linear predictor in~\cite{Ays09}. The algorithm we analyze
belongs to the general framework presented therein. Although the
algorithmic framework in~\cite{Ays09} allows for multi-tap linear
predictors, the analysis focuses entirely on one-tap prediction. Since
one-tap prediction uses only the current state-value (and the
output of neighbourhood averaging), the procedure is equivalent to
modification of the memoryless consensus weight matrix. As
such, the convergence rate improvement cannot be better than that achieved
by optimizing the weight matrix as in~\cite{Xiao04,Xiao05,Boyd06}. Aysal
et al.~also present numerical simulations for acceleration involving
multi-tap predictors, which showed much greater improvement in
convergence rate. However, they provided no method to choose or
initialize the algorithmic parameters, so it was impossible to
implement the algorithm in practice. There was no
theoretical analysis demonstrating that the predictive acceleration
procedure could consistently outperform memoryless
consensus and no characterization of the improvement.

An extreme approach to consensus acceleration is the methodology
proposed in~\cite{Sundaram07}.  Based on the notion of observability
in linear systems, the algorithm achieves consensus in a finite number
of iterations.  Each node records the entire history of values
$\{x_i(t)\}_{t=0}^T$, and after enough iterations, inverts this
history to recover the network average.  In order to carry out the
inversion, each node needs to know a topology-dependent set of
weights.  This leads to complicated initialization procedures for
determining these weights. Another drawback is that the memory
required at each node grows with the network size.

\subsection{Summary of Contributions}
\label{subsection::Summary of Contributions}

We analyze a simple, scalable and efficient framework for accelerating
distributed average consensus.  This involves the convex combination
of a neighborhood averaging and a local linear prediction. We
demonstrate theoretically that a simple two-tap linear predictor is
sufficient to achieve dramatic improvements in the convergence
rate. For this two-tap case, we provide an analytical solution for the
optimal mixing parameter and characterize the achieved improvement in
convergence rate.  We show that the performance gain grows with
increasing network size at a rate that depends on the (expected)
spectral gap of the original weight matrix\footnote{The expectation is
  appropriate for families of random graphs, and is taken over the set
  of random graphs for a specified number of nodes. For deterministic
  topologies (e.g., grid, chain) the same result applies without
  expectation.}.  As concrete examples, we show that for a chain
topology on $N$ nodes, the proposed method achieves a factor of $N$
improvement over memoryless consensus, and for a two-dimensional grid,
a factor of $\sqrt{N}$ improvement, in terms of the number of
iterations required to reach a prescribed level of accuracy. We report
the results of numerical experiments comparing our proposed algorithm
with standard memoryless consensus, the polynomial filter approach
of~\cite{Kok09} and finite-time consensus~\cite{Sundaram07}. The
proposed algorithm converges much more rapidly than memoryless
consensus, outperforms the polynomial filtering approach of~\cite{Kok09}, and achieves
performance comparable to finite-time consensus for random geometric
graph topologies. We also present a novel, efficient approach for
initialization of the accelerated algorithm. The initialization
overhead is much less than that of other acceleration methods,
rendering the scheme more practical for implementation.

\subsection{Paper Organization}
\label{subsection::Paper Organization}

The remainder of this paper is structured as follows.
Section~\ref{section::Problem Formulation} introduces the distributed
average consensus framework and outlines the linear prediction-based
acceleration methodology.  Section~\ref{section::MainResults} provides
the main results, including the optimal value of the mixing parameter
for the two-tap predictor, an analysis of convergence rate and
processing gain, and a practical heuristic for efficient distributed
initalization.  We report the results of numerical experiments in
Section~\ref{section::Numerical Examples}, and provide proofs of the
main results together with accompanying discussion in
Section~\ref{section::Proofs}. Section~\ref{section::Concluding
  Remarks} concludes the paper.

\section{Problem Formulation}
\label{section::Problem Formulation}

We assume that a network of $N$ nodes is given, and that the
communication topology is specified in terms of a collection of
neighborhoods of each node: $\N_i \subseteq \{1,\dots,N\}$ is the
set of nodes with whom node $i$ communicates directly.  For $j \in
\N_i$, we will also say that there is an edge between $i$ and $j$,
and assume that connectivity is symmetric; i.e., $j \in \N_i$
implies that $i \in \N_j$. The cardinality of $\N_i$, $d_i =
|\N_i|$, is called the degree of node $i$. We assume that the
network is connected, meaning that there is a path (a sequence of
adjacent edges) connecting every pair of nodes.

Initially, each node $i=1,\dots,N$ has a scalar value $x_i(0) \in
\Re$, and the goal is to develop a distributed algorithm such that
every node computes $\bar{x}(0) = \frac{1}{N}\sum_{i=1}^N x_i(0)$.
Previous studies (see, e.g., \cite{Tsitsiklis84} or \cite{Xiao04})
have considered linear updates of the form
\begin{equation}
x_{i}(t+1) = W_{ii}x_{i}(t) +
\sum_{j\in\mathcal{N}_{i}}W_{ij}x_{j}(t),
\end{equation}
where $\sum_{j} W_{ij} =1$, and $W_{i,j} \neq 0$ only if $j \in
\N_i$. Stacking the values $x_1(t), \dots, x_N(t)$ into a column
vector, one network iteration of the algorithm is succinctly
expressed as the linear recursion $\bx(t+1) = \bW \bx(t)$.  Let $\1$
denote the vector of all ones.  For this basic setup, Xiao and
Boyd~\cite{Xiao04} have shown that necessary and sufficient
conditions on $\bW$ which ensure convergence to the average
consensus, $\bar{\bx}(0) = \bar{x}(0) \1$, are
\begin{equation} \label{eqn:W_asym_cond}
\mathbf{W}\mathbf{1} = \mathbf{1}, \quad \mathbf{1}^{T}\mathbf{W}
=\mathbf{1}^{T}, \quad \rho(\mathbf{W} - \J) < 1,
\end{equation}
where $\J$ is the averaging matrix, $\J = \frac{1}{N} \1 \1^T$, and $\rho(\mathbf{A})$ denotes the spectral radius of a matrix $\mathbf{A}$:
\begin{equation} \label{eqn:rho}
\rho(\mathbf{A}) \triangleq
\max_{i}\{|\lambda_{i}|:i=1,2,\ldots,N\},
\end{equation}
where $\{\lambda_{i}\}_{i=1}^{N}$ denote the eigenvalues of
$\mathbf{A}$.  Algorithms have been identified for locally
generating weight matrices that satisfy the required convergence
conditions if the underlying graph is connected, e.g.,
Maximum--degree and Metropolis--Hastings
weights~\cite{Xiao07,Xiao05}.

Empirical evidence suggests that the convergence of the algorithm can
be significantly improved by using local
memory~\cite{Ays09,Joh08,Kok09}. The idea is to exploit smooth convergence
of the algorithm, using current and past values to predict the future
trajectory.  In this fashion, the algorithm achieves faster
convergence by bypassing intermediate states. Each update
becomes a weighted mixture of a prediction and a neighborhood
averaging, but the mixture weights must be chosen carefully to
ensure convergence.

The simplest case of local memory is two taps (a single tap is
equivalent to storing only the current value, as in standard
distributed averaging), and this is the case we consider in this
paper. The primary goal of this paper is to prove that local memory
can always be used to improve the convergence rate and show that the
improvement is dramatic; it is thus sufficient to examine the
simplest case. For two taps of memory, prediction at node $i$ is
based on the previous state value $x_i(t-1)$, the current value
$x_i(t)$, and the value achieved by one application of the original
averaging matrix, i.e. $x_i^{\mathrm{W}}(t+1) = W_{ii}x_i(t) +
\sum_{j \in \mathcal{N}_i}W_{ij}x_j(t)$. The state-update equations at a node
become a combination of the predictor and the value derived by
application of the consensus weight matrix (this is easily extended for predictors with longer memories; see~\cite{Joh08,Ays09}). In the
two-tap memory case, we have:
\begin{subequations} \label{eqn:pst}
\begin{align}
x_i(t+1) &= \alpha x_i^{\mathrm{P}}(t+1) + (1 - \alpha)x_i^{\mathrm{W}}(t+1) \label{eqn:pst:x_i} \\
x_i^{\mathrm{W}}(t+1) &= W_{ii}x_i(t) +
\sum_{j \in \mathcal{N}_i}W_{ij}x_j(t)  \label{eqn:pst:x_w}\\
x_i^{\mathrm{P}}(t+1) &= \theta_3 x_i^{\mathrm{W}}(t+1) + \theta_2 x_i(t) + \theta_1 x_i(t-1). \label{eqn:pst:x_p}
\end{align}
\end{subequations}
Here $\bm{\theta} = [\theta_1,\theta_2,\theta_3]$ is the vector
of predictor coefficients.

The network-wide equations can then be expressed in matrix form by
defining
\begin{align}
\bW_{3}[\alpha] &\triangleq (1-\alpha+\alpha\theta_3)\bW + \alpha\theta_{2}\bI, \label{eqn:Wmalpha}\\
\bX(t) &\triangleq [\bx(t)^T,\bx(t-1)^T]^T, \label{eqn:X}
\end{align}
where $\I$ is the identity matrix of the appropriate size, $\bX(t)$
is the memory vector, and
\begin{equation} \label{eqn:operator_phi_M}
\bPhi_{3}[\alpha] \triangleq \left[ \begin{matrix}
\bW_{3}[\alpha] & \alpha\theta_{1}\bI \\
\bI & \bm{0}
\end{matrix} \right].
\end{equation}
Each block of the above matrix has dimensions $N \times N$.  We also define $\bx(-1) = \bx(0)$ so that $\bX(0) = [\bx(0)^T \bx(0)^T]$.  The update
equation is then simply $\bX(t+1)=\bPhi_{3}[\alpha] \bX(t)$.

\section{Main Results}
\label{section::MainResults}

This section presents the main results of the paper. Proofs and more
detailed discussion are deferred to Section~\ref{section::Proofs}.
We first present in Section~\ref{ssection::OptMixResults} a
discussion of how to optimize the two-tap memory predictive
consensus algorithm with respect to the network topology. The main
contribution is an analytical expression for the mixing parameter
$\alpha$ that achieves the minimum limiting convergence time (a
concept defined below). This analytical expression involves only the
second-largest eigenvalue of the original weight matrix $\bW$. In
Section~\ref{ssection::initialization}, we describe an efficient
distributed algorithm for estimating the second-largest eigenvalue.
This means that there is only a relatively small overhead in
initializing the predictive consensus algorithm with a very accurate
approximation to the optimal mixing parameter.

Section~\ref{ssection::conv_rate_proposed} presents an analysis of
the convergence rate of the two-tap memory predictor-based consensus
algorithm when the optimal mixing parameter is used. We show how
incorporating prediction affects the spectral radius, which governs
asymptotic convergence behaviour. Our result provides a bound on how
the spectral radius scales as the number of nodes in the network is
increased. We discuss how this bound can be used to develop
guidelines for selecting asymptotically optimal prediction
parameters $\bm{\theta}$.  The second set of results on convergence
time, presented in Section~\ref{ssection::processing_gain},
characterizes a processing gain metric. This metric measures the
improvement in asymptotic convergence rate achieved by an
accelerated consensus algorithm (relative to the convergence rate
achieved by standard distributed averaging using the original weight
matrix).

\subsection{Optimal Mixing Parameter}
\label{ssection::OptMixResults}

The mixing parameter $\alpha$ determines the influence of the
standard one-step consensus iteration relative to the predictor in
\eqref{eqn:pst:x_i}. We assume a foundational weight matrix, $\bW$,
has been specified, and proceed to determine the optimal mixing
parameter $\alpha$ with respect to $\bW$.  Before deriving an
expression for the optimal $\alpha$, it is necessary to specify what
``optimal'' means. Our goal is to minimize convergence time, but it
is important to identify how we measure convergence time.

Xiao and Boyd~\cite{Xiao04} show that selecting
weights $\bW$ to minimize the spectral radius $\rho(\bW - \bJ)$ (while
respecting the network topology constraints) leads to the optimal
convergence rate for standard distributed averaging.  In particular,
the spectral radius is the worst-case asymptotic convergence rate,
\begin{equation}
\rho(\bW - \bJ) = \sup_{\bx(0) \ne \bar{\bx}(0)} \lim_{t \rightarrow
\infty} \left(\frac{\|\bx(t) - \bar{\bx}(0)\|}{\|\bx(0) -
\bar{\bx}(0)\|}\right)^{1/t}.
\end{equation}
Maximizing asymptotic convergence rate is equivalent to
minimizing asymptotic convergence time,
\begin{equation}
\tau_{\scriptsize \mbox{asym}} \triangleq \frac{1}{\log(\rho(\bW - \bJ)^{-1})},
\end{equation}
which, asymptotically, corresponds to the number of iterations
required to reduce the error $\|\bx(t) - \bar{\bx}(0)\|$ by a factor
of $e^{-1}$ \cite{Xiao04}. An alternative metric is the
\emph{convergence time}, the time required to achieve the prescribed
level of accuracy $\varepsilon$ for any non-trivial
initialization~\cite{Ols09}:
\begin{align} \label{eqn:t_ave}
T_{\con}(\W, \varepsilon) = \inf_{\tau \geq 0} \left\{ \tau : ||
\bx(t) - \bar{\bx}(0) ||_2 \leq \varepsilon || \bx(0) - \bar{\bx}(0)
||_2 \quad \forall\ t \geq \tau , \quad \forall\ \bx(0)-\bar{\bx}(0)
\neq \bZ\right\},
\end{align}
In the case where $\bW$ is symmetric, $\rho(\W-\J)$ also defines the
convergence time~\cite{Oresh09}. The update matrix we propose,
(\ref{eqn:operator_phi_M}), is not symmetric and it may not even be
contracting. For such matrices, and the spectral radius
$\rho(\W-\J)$ cannot, in general, be used to specify an upper bound
on convergence time. We can, however, establish a result for the
{\em limiting} $\varepsilon$-convergence time, which is the
convergence time for asymptotically small $\varepsilon$.
Specifically, in Section~\ref{ssection::limiting_averaging_time} we
show that for matrices of the form \eqref{eqn:operator_phi_M},
\begin{align}
\lim\limits_{\varepsilon \rightarrow 0}
\frac{T_{\con}(\bPhi_{3}[\alpha], \varepsilon)}{\log
\varepsilon^{-1}} = \frac{1}{\log \rho(\bPhi_{3}[\alpha] -
\J)^{-1}}.
\end{align}
According to this result, the convergence time required to approach
the average within $\varepsilon$-accuracy grows at the rate $1/\log
\rho(\bPhi_{3}[\alpha] - \J)^{-1}$ as $\varepsilon \rightarrow 0$.
Minimizing the spectral radius is thus a natural optimality
criterion. The following theorem establishes the optimal setting of
$\alpha$ for a given weight matrix $\bW$, as a function of
$\lambda_2(\bW)$, the second largest eigenvalue of $\bW$.

\begin{theorem}[Optimal mixing parameter] \label{lem:optimization}
  Suppose $\bW \in \mathbb{R}^{N\times N}$ is a symmetric weight matrix
  satisfying convergence conditions~\eqref{eqn:W_asym_cond} and $|\lambda_N(\bW)| \leq
  \lambda_2(\bW)$, where the eigenvalues $\lambda_1(\bW)=1,\lambda_2(\bW),\dots,\lambda_N(\bW)$
  are labelled in decreasing order. Suppose further that $\theta_{3}+\theta_{2}+\theta_{1}=1$ and $\theta_{3} \geq
  1$, $\theta_{2} \geq 0$. Then the solution of the optimization problem
\begin{align} \label{eqn:opt_prob}
\alpha^{\star} = \arg \min\limits_{\alpha} \rho(\bPhi_{3}[\alpha] -
\bJ)
\end{align}
is given by the following:
\begin{align}
\label{eq:optalpha}
\alpha^{\star} &= \frac{-((\theta_{3}-1) \lambda_2(\bW)^2 +
\theta_{2} \lambda_2(\bW) + 2\theta_{1}) - 2\sqrt{\theta_{1}^2 +
\theta_{1}\lambda_2(\bW)\left(\theta_{2}+(\theta_{3}-1)\lambda_2(\bW)
\right)}}{\left(\theta_{2} + (\theta_{3}-1)\lambda_2(\bW)\right)^2 }
\end{align}
\end{theorem}

A brief discussion of the conditions of this theorem is warranted.
The conditions on the predictor weights are technical conditions
that ensure convergence is achieved. Two factors motivate our belief
that these are not overly-restricting. First, these conditions are
satisfied if we employ the least-squares predictor weights design
strategy. Aysal et al.~\cite{Ays09} describe a method for choosing
the predictor coefficients $\bm{\theta}$ based on least-squares
predictor design. For the two-tap memory case, the predictor
coefficients are identified as $\bm{\theta} = \A^{\dag T} \B$, where
\begin{equation}\label{eqn:At_plus_t_pk} \A \triangleq
\left[
\begin{matrix}
-2  & -1 & 0\\
1  & 1 & 1
\end{matrix}
\right]^T\,\,,
\end{equation}
$\B\triangleq[1,1]^T$, and $\A^{\dag}$ is the Moore-Penrose
pseudoinverse of $\bA$. This choice of predictor coefficients
satisfies the technical conditions on $\bm{\theta}$ in
Theorem~\ref{lem:optimization} above ($\theta_1 + \theta_2 +
\theta_3 = 1$ and $\theta_3 \geq 1, \theta_2 \geq 0$). Second, in
Section~\ref{ssection::conv_rate_proposed} we show that the choice
of weights does not have a significant effect on the convergence
properties, and asymptotically optimal weights also satisfy
conditions on $\bm{\theta}$ in Theorem~\ref{lem:optimization}.

The condition on the weight matrix, $|\lambda_N(\bW)| \leq
\lambda_2(\bW)$, significantly reduces the complexity of the proof.
Most distributed algorithms for constructing weight matrices (e.g.,
Metropolis-Hastings (MH) or max-degree) lead to $\bW$ that satisfy
the condition, but they are not guaranteed to do so. We can ensure
that the condition is satisfied by applying a completely local
adjustment to any weight matrix. The mapping $\bW \mapsto 1/2(\bI +
\bW)$ transforms any stochastic matrix $\bW$ into a stochastic
matrix with all positive eigenvalues~\cite{Boyd06}; this mapping can
be carried out locally, without any knowledge of the global
properties of $\bW$, and without affecting the order-wise asymptotic
convergence rate as $N\rightarrow \infty$.

\subsection{Convergence Rate Analysis}
\label{ssection::conv_rate_proposed}

We begin with our main result for the convergence rate of two-tap
predictor-based accelerated consensus.
Theorem~\ref{th:conv_rate_bound} indicates how the spectral radius
of the accelerated operator $\bPhi_3[\alpha]$ is related to the
spectral radius of the foundational weight matrix $\bW$. Since the
limiting $\varepsilon$-convergence time is governed by the spectral
radius, this relationship characterizes the improvement in
convergence rate.

\begin{theorem} [Convergence rate] \label{th:conv_rate_bound}
Suppose the assumptions of Theorem~\ref{lem:optimization} hold.
Suppose further that the original matrix $\bW$ satisfies $\rho(\bW -
\J) \leq 1 - \Psi(N)$ for some function $\Psi: \mathbb{N}
\rightarrow (0, 1)$ of the network size $N$. Then the matrix
$\bPhi_3[\alpha^{\star}]$ satisfies $\rho(\bPhi_3[\alpha^{\star}] -
\J) \leq 1 - \sqrt{\Psi(N)}$.
\end{theorem}

In order to explore how fast the spectral radius,
$\rho(\bPhi_3[\alpha^{\star}]-\J) = \sqrt{-\alpha^{\star}\theta_{1}}$,
(see Section~\ref{ssection::conv_rate_proof} for details) goes to
one as $N\rightarrow \infty$, we can take its asymptotic Taylor
series expansion:
\begin{align} \label{eqn:asym_acc_bound}
\rho(\bPhi_3[\alpha^*]-\J) &= 1 -
\sqrt{\frac{2(\theta_{3}-1)+\theta_{2}}{\theta_{3}-1+\theta_{2}}}\sqrt{\Psi(N)}
+ \mathcal{O}(\Psi(N)).
\end{align}
From this expression, we see that the bound presented in
Theorem~\ref{th:conv_rate_bound} correctly captures the convergence
rate of the accelerated consensus algorithm. Alternatively, leaving
only two terms in the expansion above, $\rho(\bPhi_3[\alpha^*]-\J) = 1 - \Omega(\sqrt{\Psi(N)})$,  we see that the bound presented is rate optimal in Landau notation.

We can also use (\ref{eqn:asym_acc_bound}) to provide guidelines for
choosing asymptotically optimal prediction parameters $\theta_{3}$
and $\theta_{2}$. In particular, it is clear that the coefficient
$\gamma(\theta_{2}, \theta_{3}) =
\sqrt{[2(\theta_{3}-1)+\theta_{2}]/[\theta_{3}-1+\theta_{2}]}$
should be maximized to minimize the spectral radius
$\rho(\bPhi_{3}[\alpha^{\star}] - \J)$. It is straightforward to
verify that setting $\theta_{2}=0$ and $\theta_{3} = 1+\epsilon$ for
any $\epsilon > 0$ satisfies the assumptions of
Theorem~\ref{lem:optimization} and also satisfies $\gamma(0,
1+\epsilon) > \gamma(\theta_{2}, 1+\epsilon)$ for any positive
$\theta_{2}$.  Since $\gamma(0, 1+\epsilon) = \sqrt{2}$ is
independent of $\epsilon$ (or $\theta_{3}$) we conclude that setting
$(\theta_1, \theta_2, \theta_3) = (-\epsilon, 0, 1+\epsilon)$
satisfies the assumptions of Theorem~\ref{lem:optimization} and
asymptotically yields the optimal limiting $\varepsilon$-convergence
time for the proposed approach, as $N \rightarrow \infty$.

\subsection{Processing Gain Analysis}
\label{ssection::processing_gain}

Next, we investigate the gain that can be obtained by using the
accelerated algorithm presented in this paper. We consider the ratio
$\tau_{\asym}(\bW) / \tau_{\asym}(\bPhi_3[\alpha^*])$ of the asymptotic
convergence time of the standard consensus algorithm using weight
matrix $\bW$ and the asymptotic convergence time of the proposed accelerated
algorithm. This ratio shows how many
times fewer iterations, asymptotically, the optimized predictor-based
algorithm must perform to reduce error by a factor of $e^{-1}$.

If the network topology is modeled as random (e.g., a sample from the
family of random geometric graphs), we adopt the expected gain $\mathcal{G}(\bW) = \E\{
\tau_{\asym}(\bW) / \tau_{\asym}(\bPhi_3[\alpha^{\star}]) \}$ as a
performance metric, where $\bPhi_3[\alpha^*]$ is implicitly constructed using the same matrix $\bW$.  The expected gain characterizes the average
improvement obtained by running the algorithm over many realizations
of the network topology. In this case the spectral radius, $\rho(\bW - \J)$,
is considered to be a random variable dependent on the particular
realization of the graph. Consequently, the expectations in the
following theorem are taken with respect to the measure induced by
the random nature of the graph.

\begin{theorem} [Expected gain] \label{th:t_lim_ratio}
Suppose the assumptions of Theorem~\ref{lem:optimization} hold.
Suppose further that the original matrix $\bW$ satisfies
$\E\{\rho(\bW - \J)\} = 1 - \Psi(N)$ for some function $\Psi:
\mathbb{N} \rightarrow (0, 1)$ of the network size $N$. Then
$\mathcal{G}(\bW) = 1/\sqrt{\Psi(N)}$.
\end{theorem}

We note that there is no loss of generality in considering the
expected gain since, in the case of a deterministic network
topology, these results will still hold (without expectations) since
they are based on the deterministic derivations in Theorems~1 and 2.

For a chain graph (path of $N$ vertices) the eigenvalues of the
Metropolis-Hastings (MH) weight matrix, $\bW_{\text{MH}}$,
constructed according to~\cite{Xiao05} ($\bW_{i,j} = 1/(1+\max(d_i,
d_j))$ if $j \in \mathcal{N}_i, i\neq j$; $\bW_{i,j} = 0$ if $j
\notin \mathcal{N}_i$; and $\bW_{i,i} = 1 - \sum_{j \in
{\mathcal{N}_i}} \bW_{i,j}$) are given by
$\lambda_{i}(\bW_{\text{MH}}) = 1/3 + 2/3\cos(\pi (i-1) / N), i = 1,
2, \ldots, N$. This is straightforward to verify using Theorem~5
in~\cite{Yueh05}. For the path graph, the weight matrix
$\bW_{\text{MH}}$ is tridiagonal and we have $\max(d_i, d_j) = 2,
\forall i,j$. Thus, in this case, $\rho(\bW_{\text{MH}} - \J) = 1/3
+ 2/3 \cos(\pi/N)$. For large enough $N$ this results in
$\rho(\bW_{\text{MH}} - \J) \approx 1 - \frac{\pi^2}{3}
\frac{1}{N^2} + \mathcal{O}(1/N^4)$. Using the same sequence of
steps used to prove Theorem~\ref{th:t_lim_ratio} above without
taking expectations, we see that for the chain topology, the
improvement in asymptotic convergence rate is asymptotically lower
bounded by $N$; i.e., $\mathcal{G}(\bW) = \Omega(N)$. Similarly, for
a network with two-dimensional grid topology, taking $\bW$ to  be
the transition matrix for a natural random walk on the grid (a minor
perturbation of the MH weights) it is known~\cite{aldousFill} that
$(1-\lambda_2(\bW))^{-1} = \Theta(N)$. Thus, for a two-dimensional
grid, the proposed algorithm leads to a gain of $\mathcal{G}(\bW) =
\Omega(N^{1/2})$.

This discussion suggests that the following result may also be useful
in characterizing the improvement in asymptotic convergence rate
obtained by using the proposed algorithm.
\begin{corollary}
Suppose that assumptions of Theorem~\ref{th:t_lim_ratio} hold and
suppose in addition that $\rho(\bW - \J) = 1 -
\Theta(\frac{1}{N^{\beta}})$ then the improvement in asymptotic
convergence rate attained by the accelerated algorithm is $\mathcal{G}(\bW) = \Omega(N^{\beta/2})$.
\end{corollary}

\subsection{Initialization Heuristic: Decentralized Estimation of
$\lambda_2(\bW)$}
\label{ssection::initialization}

Under our assumptions, the optimal value of the mixing parameter
depends only on the values of predictor coefficients and the second
largest eigenvalue of initial matrix $\bW$. In this section we
discuss a decentralized procedure for estimating $\lambda_2(\bW)$.
Since we assume the predictor weights, $\bm{\theta}$, and weight
matrix $\bW$ are fixed and specified, this is the only parameter
that remains to be identified for a fully decentralized
implementation of the algorithm. Estimation of $\lambda_2(\bW)$ is a
straightforward exercise if we employ the method of decentralized
orthogonal iterations (DOI) proposed for distributed spectral
analysis in~\cite{Kempe04} and refined for distributed optimization
applications in~\cite{Boyd06}.

Algorithm~\ref{alg:eig_est_sup} presents the proposed specialized
and streamlined version of DOI, which is only used to calculate the
second largest eigenvalue of the consensus update matrix $\bW$. Our
underlying assumptions in Algorithm~\ref{alg:eig_est_sup} are those
of Theorem~\ref{lem:optimization}, in which case we have
$\lambda_2(\bW) = \rho(\bW-\bJ)$. The eigenvalue shifting technique
discussed after Theorem~\ref{lem:optimization} can be employed
whenever assumption $|\lambda_N(\bW)| \leq \lambda_2(\bW)$ does not
hold. The main idea of DOI, is to repeatedly apply $\bW$ to a random
vector $\bv_0$, with periodic normalization and subtraction of the
estimate of the mean, until $\bv_{K} = \bW^K \bv_0$ converges to the
second-largest eigenvector of $\bW$. Then, estimate the
second-largest eigenvalue by calculating
$||\bW\bv_{K}||/||\bv_{K}||$ for a valid matrix norm $\| \cdot \|$.
Previous algorithms for DOI~\cite{Kempe04,Boyd06} have normalized in
step 6 by the $\ell_2$ norm of $\bv_k$, estimated by $K$ iterations
of consensus, and step 9 previously required an additional $K$
iterations to calculate $\|\bW\bv_K\|_{2}$ and $\|\bv_K\|_{2}$. In
addition, because the initial random vectors
in~\cite{Kempe04,Boyd06} are not zero-mean, these algorithms must
apply additional consensus operations to eliminate the bias
(otherwise $\bv_K$ converges to $\1$). Previous algorithms thus have
$\mathcal{O}(K^2)$ complexity, where $K$ is the topology-dependent
number of consensus iterations needed to achieve accurate
convergence to the average value.  For example, for a random
geometric graph, one typically needs $K \propto N$.

\begin{algorithm}[ht] \label{alg:eig_est_sup}

\SetKwComment{Comment}{}{}

Choose random vector $\bv$ \;
Set $\bv_0 = \bW\bv -\bv$ \Comment*[l]{Generate zero-mean random vector}

\For{$k=1$ \textrm{to} $K$ } {

$\bv_{k} = \bW \bv_{k-1}$ \Comment*[I]{Apply $\bW$ to converge to second-largest eigenvector}

\If{$k \mod L = 0$}{
$\bv_{k} = \bv_{k} / || \bv_{k} ||_{\infty}$ \Comment*[l]{Normalize by supremum norm every $L$ iterations}
}
}%End for t

Let $\widehat{\lambda_{2}}(\bW) = \|\bW \bv_{K}\|_{\infty} /
\|\bv_{K}\|_{\infty}$ \;

\caption{Spectral radius estimation (Input: foundational weight matrix $\bW$)}
\end{algorithm}

The main innovations of Algorithm~\ref{alg:eig_est_sup} are in line
2, which ensures that the initial random vector is zero mean, in
line 6, where normalization is done (after every $L$ applications of
the consensus update) using the supremum norm, and line 9, where the
supremum norm is also used in lieu of the $\ell_2$ norm\footnote{We
have not observed any penalty for using the $\ell_\infty$ norm in
our experiments. This observation is supported by the theoretical
equivalence of $\ell_p$ norms in the consensus
framework~\cite{Ols09}.} (based on Gelfand's
formula~\cite{hornJohnson} we have $\lim_{K \rightarrow \infty}
\|\bW \bv_{K} \|_{\infty} / \| \bv_{K} \|_{\infty} = \rho(\W - \J)
$). The maximum entry of the vector $\bv_K$ can be calculated using
a maximum consensus algorithm, wherein every node updates its value
with the maximum of its immediate neighbours: $\x_i(t) = \max_{j
  \in \mathcal{N}_i} \x_j(t-1)$. Maximum consensus requires at most
$N$ iterations to converge for any topology; more precisely it
requires a number of iterations equal to the diameter, $D$, of the
underlying graph, which is often much less than $N$ (and much less
than $K$). Equally importantly, maximum consensus achieves perfect
agreement. In the algorithms of~\cite{Kempe04,Boyd06} each node
normalizes by a slightly different value (there are residual errors
in the consensus procedure). In Algorithm 1, all nodes normalize by
the same value, and this leads to much better estimation accuracy.
Taken together, these innovations lead to an algorithm that is only
$\mathcal{O}(K)$ (with the appropriate choice of $L$). In
particular, the complexity of Algorithm~\ref{alg:eig_est_sup} is
clearly $\mathcal{O}(K + DK/L + D)$. Choosing $L \propto D$
(assuming that $\lambda_2(\W)^D \gg \Delta$, where $\Delta$ is
machine precision) we obtain an $\mathcal{O}(K)$ algorithm. The
proposed initialization algorithm has significantly smaller
computation/communication complexity than the initialization
algorithm proposed for the distributed computation of optimal matrix
in~\cite{Boyd06}.

\section{Numerical Experiments and Discussion}
\label{section::Numerical Examples}

This section presents simulation results for two scenarios. In the
first simulation scenario, network topologies are drawn from the
family of random geometric graphs of $N$ nodes~\cite{gupta00}.  In
this model, $N$ nodes are randomly assigned coordinates in the unit
square, and links exist between nodes that are at most a distance
$\sqrt{2\log{N}/N}$. (This scaling law for the connectivity radius
guarantees the network is connected with high
probability~\cite{gupta00}.) Two models for the initial node
measurements, $\bx(0)$, are considered.  In the ``Slope'' model, the
initial value $x_i(0)$ at node $i$ is just the sum of its
coordinates in the unit square.  In the ``Spike'' model, all nodes
are initialized to 0, except for one randomly chosen node whose
initial value is set to one. All simulation results are generated
based on $300$ trials (a different random graph and node
initialization is generated for each trial). The initial values are
normalized so that the initial variance of node values is equal to
1.  The second simulation scenario is for the $N$-node chain
topology.  Intuitively, this network configuration constitutes one
of the most challenging topologies for distributed averaging
algorithms since the chain has the longest diameter and weakest
connectivity of all graphs on $N$.  For this topology, we adopt
analogous versions of the ``Slope" and ``Spike" initializations to
those described above; for the ``Slope", $x_i(0) = i/N$, and for the
``Spike", we average over all locations of the one.

We run the algorithm $N$ times with different initializations of the
eigenvalue estimation algorithm to investigate the effects of
initializing $\alpha^{\star}$ with an imperfect estimate of
$\lambda_2(\bW)$.  In simulations involving the calculation of
convergence time we have fixed the required accuracy of
computations, $\varepsilon$, at the level $-100$ dB (i.e., a
relative error of $1\times 10^{-5}$). For predictor parameters, we
use $(\theta_1, \theta_2, \theta_3) = (-\epsilon, 0, 1+\epsilon)$,
$\epsilon = 1/2$, as these were shown to be asymptotically optimal
in Section~\ref{ssection::conv_rate_proposed}.

We compare our algorithm with two memoryless approaches, the
Metropolis-Hastings (MH) weight matrix, and the optimal weight
matrix of Xiao and Boyd~\cite{Xiao04}~\footnote{To determine the
optimal weight matrix and optimal polynomial filter weights we used
\texttt{CVX}, a package for specifying and solving convex programs
\cite{BoydCVX}.}.  MH weights are attractive because they can be
calculated by each node simply using knowledge of its own degree and
its neighbors' degrees. We also compare to two approaches from the
literature that also make use of memory at each node to improve the
rate of convergence: polynomial filtering~\cite{Kok09}, and
finite-time consensus~\cite{Sundaram07}.

\begin{figure}[t]
\centering \subfigure[Slope initialization]{
\label{fig:feat_space:sim} %% label for first subfigure
\includegraphics[width = 8cm]{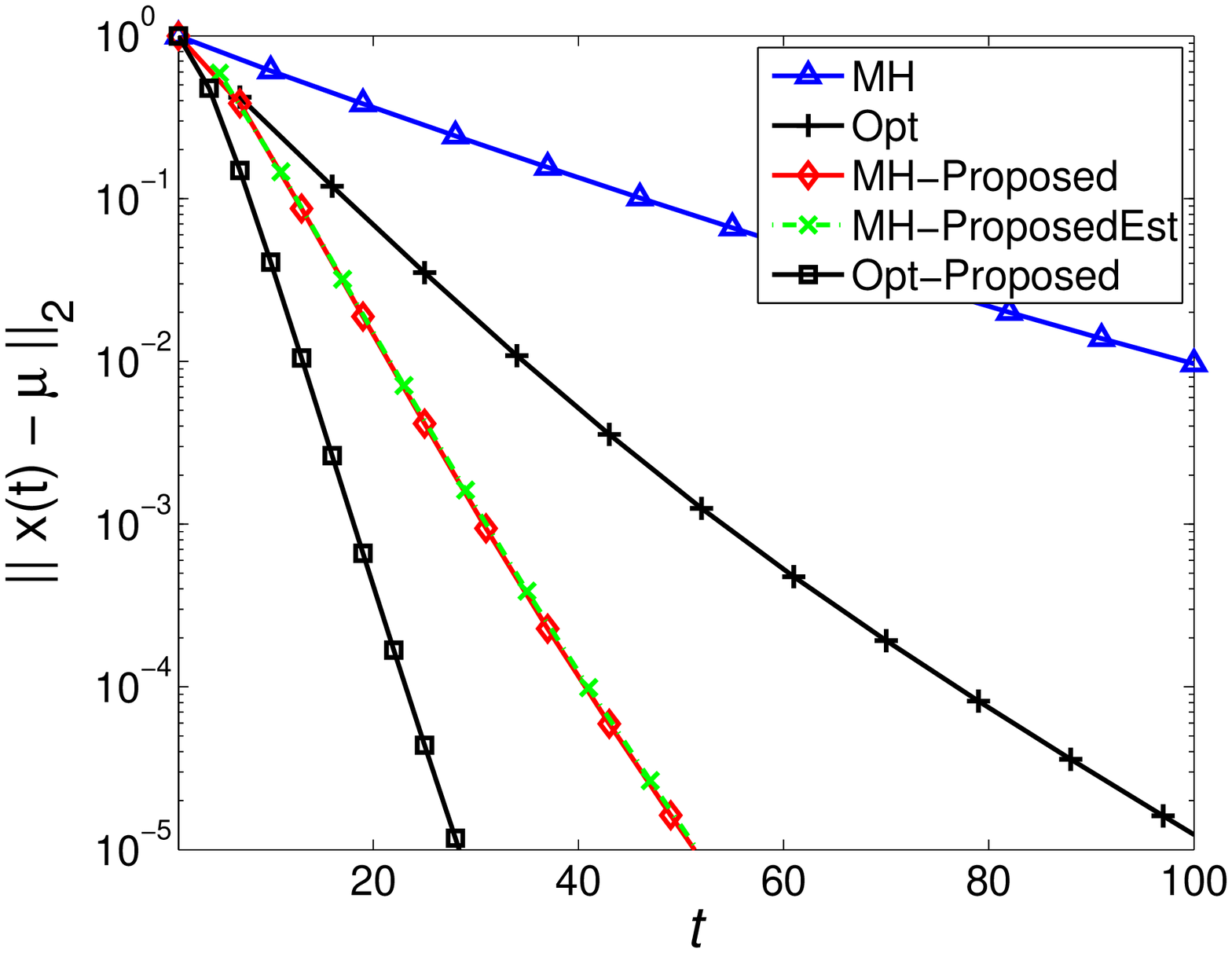}}
 \subfigure[Spike initialization]{ % \hfill
\label{fig:feat_space:real} %% label for second subfigure
\includegraphics[width = 8cm]{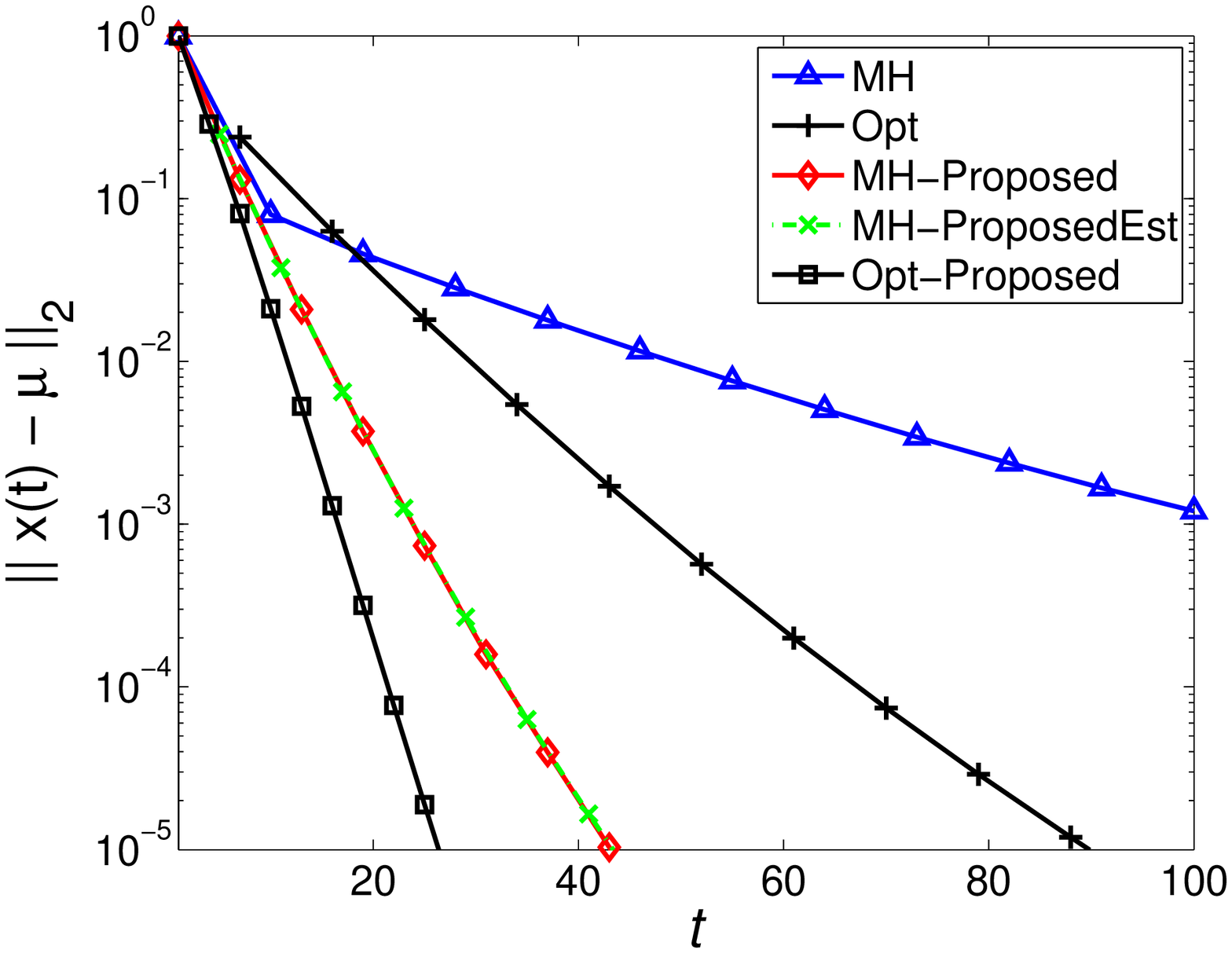}}
\caption{MSE vs.~iterations for 200-node random geometric graphs.
The algorithms compared are: optimal weight matrix of Xiao and
Boyd~\cite{Xiao04} (Opt): $+$; MH weights (MH): $\triangle$;
proposed method with oracle $\lambda_{2}(\bW)$ and MH matrix
(MH-Proposed): $\diamond$; proposed with decentralized estimate of
$\lambda_{2}(\bW)$ (MH-ProposedEst): $\times$; accelerated
consensus, with oracle $\lambda_{2}(\bW)$ and optimal matrix
(Opt-Proposed): $\square$. (a) Slope initialization. (b) Spike
initialization. }
\label{fig:mse_rg_init} %% label for entire figure*
\end{figure}

We first plot the MSE decay curves as a function of the number of
consensus iterations $t$ for network size $N=200$, RGG topology and
different initializations. Figure~\ref{fig:mse_rg_init} compares the
performance of the proposed algorithm with the algorithms using the
MH or the optimal weight matrix of Xiao and Boyd~\cite{Xiao04}.  It
can be seen that our decentralized initialization scheme does not
have a major influence on the performance of our approach, as the
method initialized using a decentralized estimate for
$\lambda_2(\bW)$ (the curve labelled MH-ProposedEst) and the method
initialized using precise knowledge of $\lambda_2(\bW)$ (labelled
MH-Proposed) coincide nearly exactly since the procedure discussed
in Section~\ref{ssection::initialization} provides a good estimate
of $\lambda_{2}(\bW)$ (to within $10^{-3}$ maximum relative error
for a 200 node RGG). It is also clear that the proposed algorithm
outperforms both the memoryless MH matrix and the optimal weight
matrix of Xiao and Boyd~\cite{Xiao04}. In this experiment we fixed
$K=2N$ and $L=10$. Note that the results in Figure 1 and all
subsequent figures do not account for initialization costs. The
initialization cost is relatively small. For the 200-node RGG it is
equal to about $3N = 600$ consensus iterations (if we bound the
diameter of the 200-node RGG by 20). If we desire a relative error
of $10^{-3}$, our algorithm gains approximately 70 iterations over
memoryless MH consensus, based on Fig.~\ref{fig:feat_space:real}.
For this desired accuracy, the initialization overhead is thus
recovered after less than 10 consensus operations.

% Figure 1

% Figure 2
\begin{figure}[t]
\centering \subfigure[Random geometric graph]{
\label{fig:feat_space:sim} %% label for first subfigure
\includegraphics[width = 8cm]{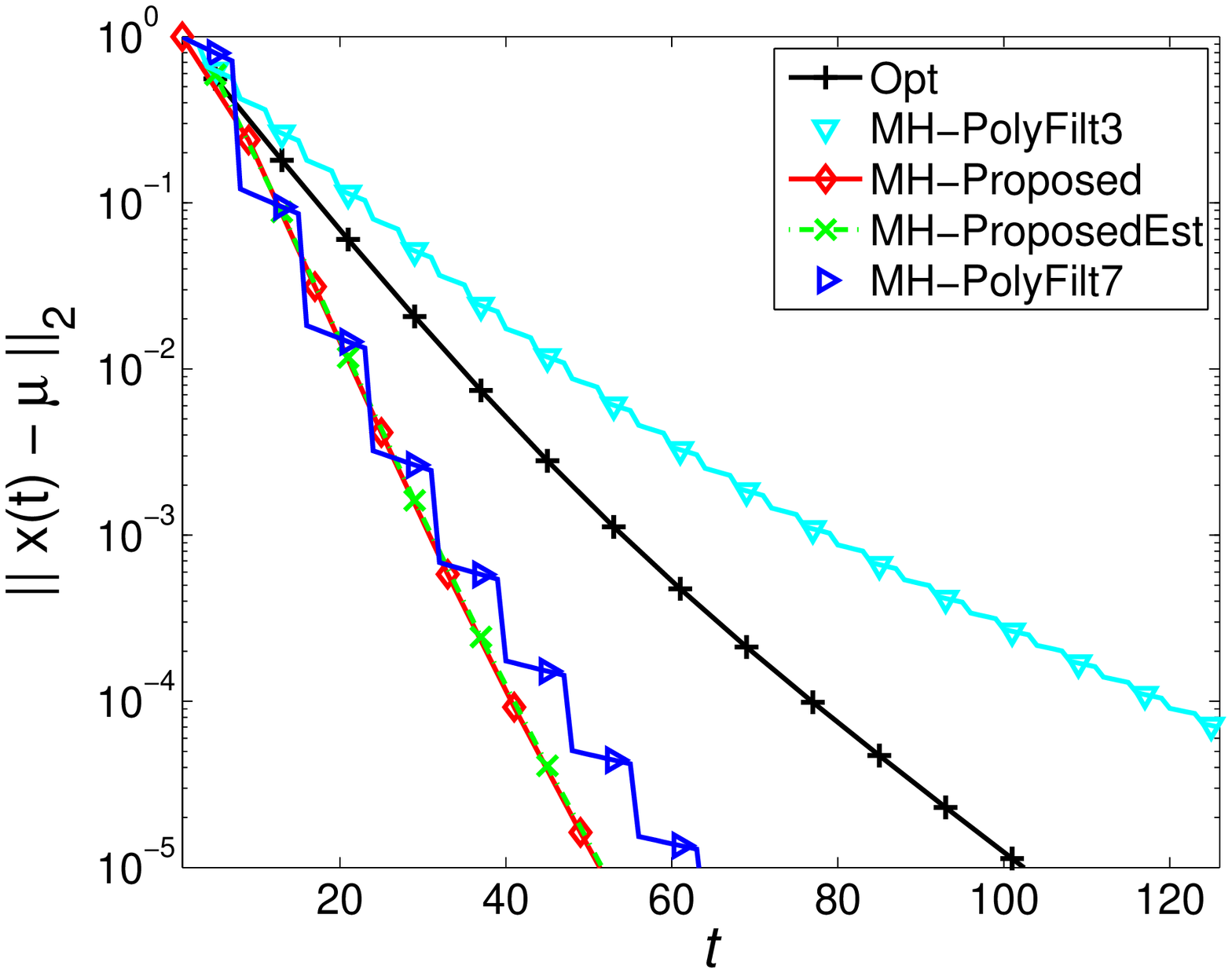}}
 \subfigure[Chain]{ % \hfill
\label{fig:feat_space:real} %% label for second subfigure
\includegraphics[width = 8cm]{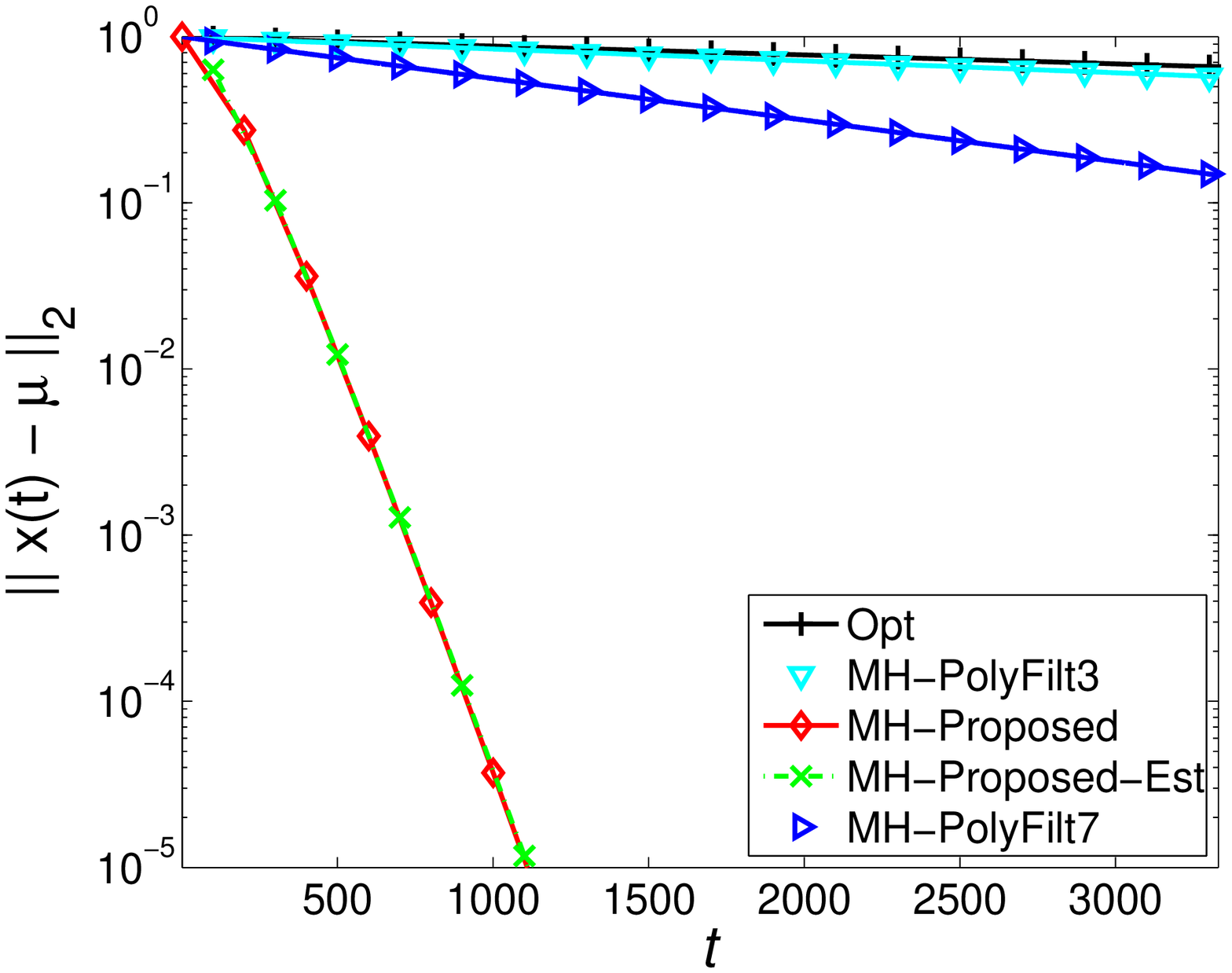}}
\caption{MSE vs.~iteration for 200-node topologies, Slope
initialization. The algorithms compared are: optimal weight matrix
of Xiao and Boyd~\cite{Xiao04} (Opt): $+$; polynomial filter with 3
taps (MH-PolyFilt3): $\bigtriangledown$ and 7 taps (MH-PolyFilt7):
$\triangleright$; proposed method with oracle $\lambda_{2}(\bW)$ and
MH matrix (MH-Proposed): $\diamond$; proposed method with
decentralized estiamte of $\lambda_{2}(\bW)$ (MH-ProposedEst):
$\times$.}
\label{fig:mse_rg_circ_acc} %% label for entire figure*
\end{figure}

% Figure 3
\begin{figure*}[t]
\centering \subfigure[ ]{
\label{fig:t_ave_rg_poly_acc_pap_true} %% label for first subfigure
\includegraphics[width = 7.5cm]{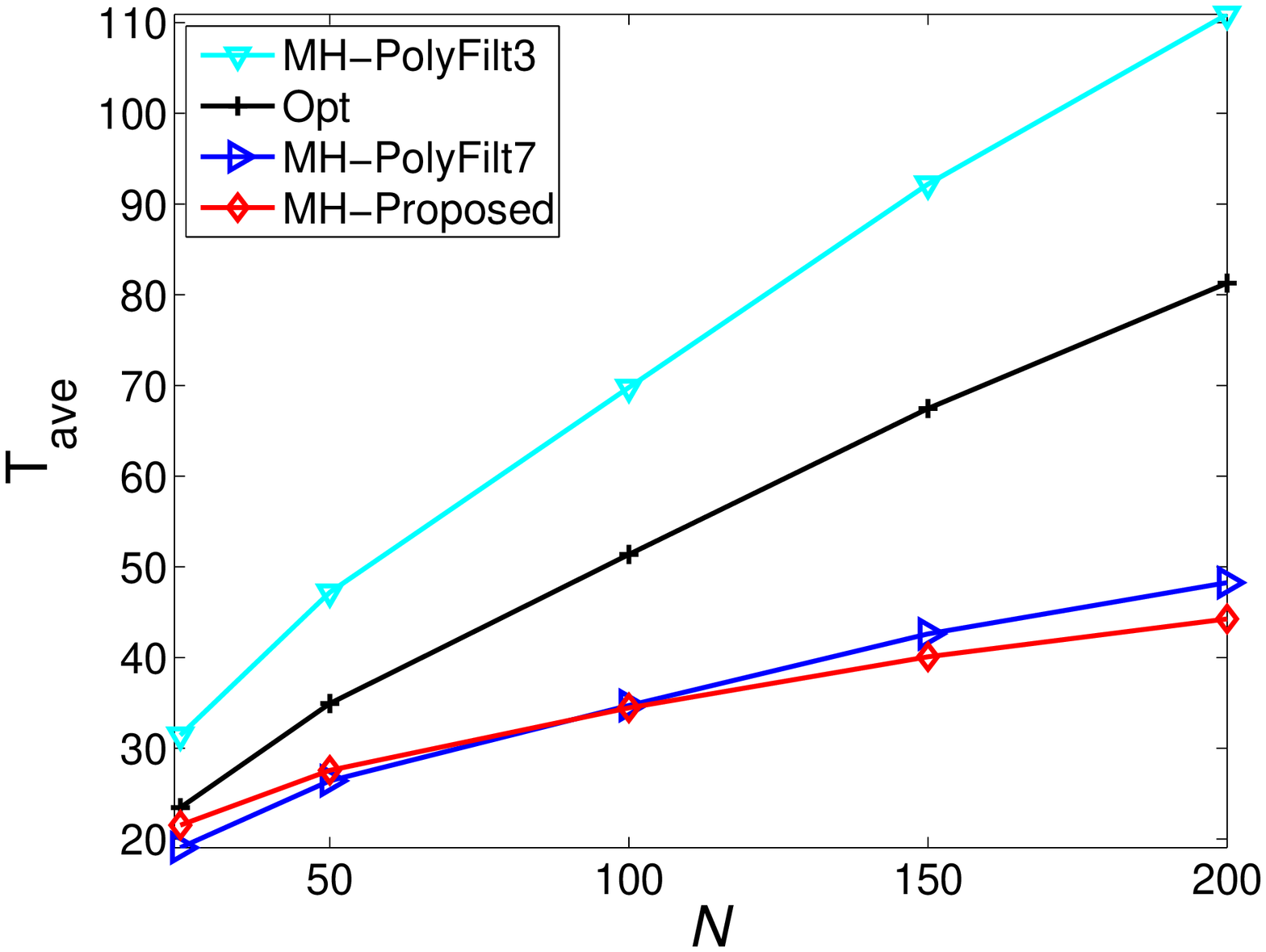}}
\hspace{1cm}
 \subfigure[ ]{ % \hfill
\label{fig:t_ave_rg_poly_acc_pap_rat} %% label for second subfigure
\includegraphics[width = 7.5cm]{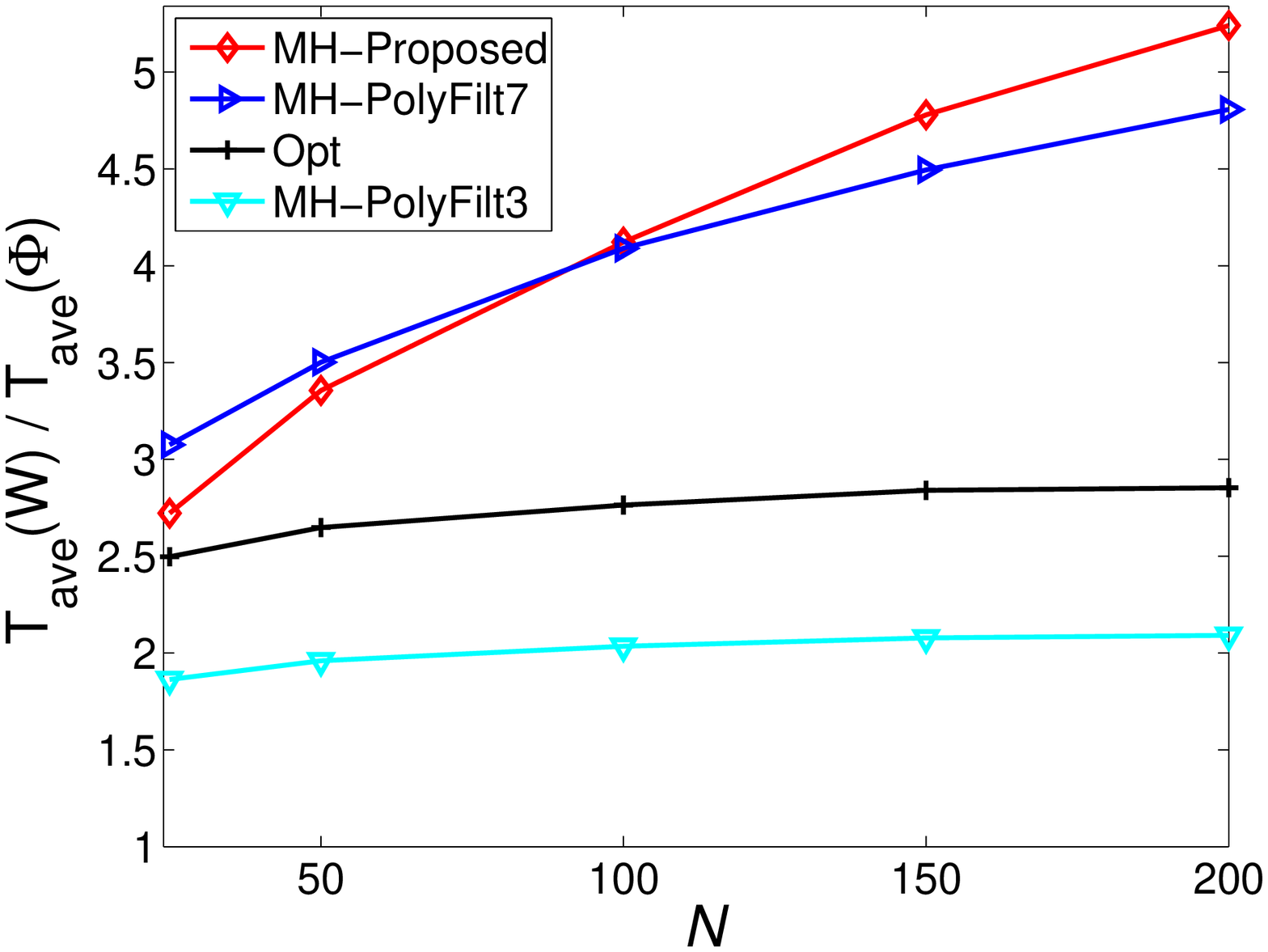}}
\caption{Averaging time characterization, random geometric graph
topologies. The algorithms compared are: optimal weight matrix of
Xiao and Boyd~\cite{Xiao04} (Opt): $+$; polynomial filter with 3
taps (MH-PolyFilt3): $\bigtriangledown$, and 7 taps (MH-PolyFilt7):
$\triangleright$; proposed method with oracle $\lambda_{2}(\bW)$ and
MH matrix (MH-Proposed): $\diamond$; proposed method with MH matrix
and decentralized estimate of $\lambda_{2}(\bW)$ (MH-ProposedEst):
$\times$. (a) Averaging time as a function of the network size. (b)
Ratio of the averaging time of the non-accelerated algorithm to that
of the associated accelerated algorithm.}
\label{fig:t_ave_rgg} %% label for entire figure*
\end{figure*}

% Figure 4
\begin{figure*}[t]
\centering \subfigure[ ]{
\label{fig:t_ave_circ_poly_acc_pap_true} %% label for first subfigure
\includegraphics[width = 7.6cm]{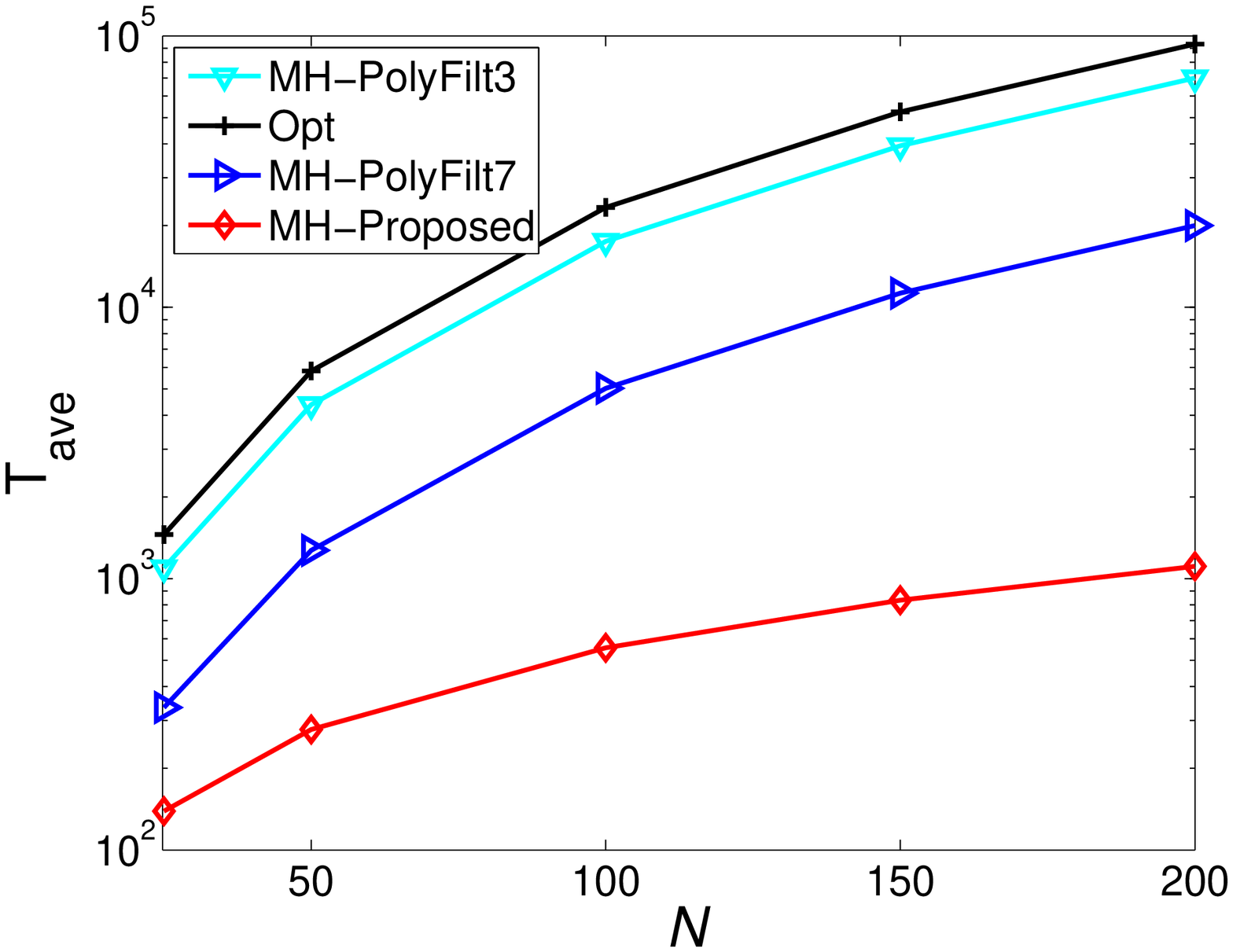}}
%\hspace{1cm}
 \subfigure[ ]{ % \hfill
\label{fig:t_ave_circ_poly_acc_pap_rat} %% label for second subfigure
\includegraphics[width = 7.6cm]{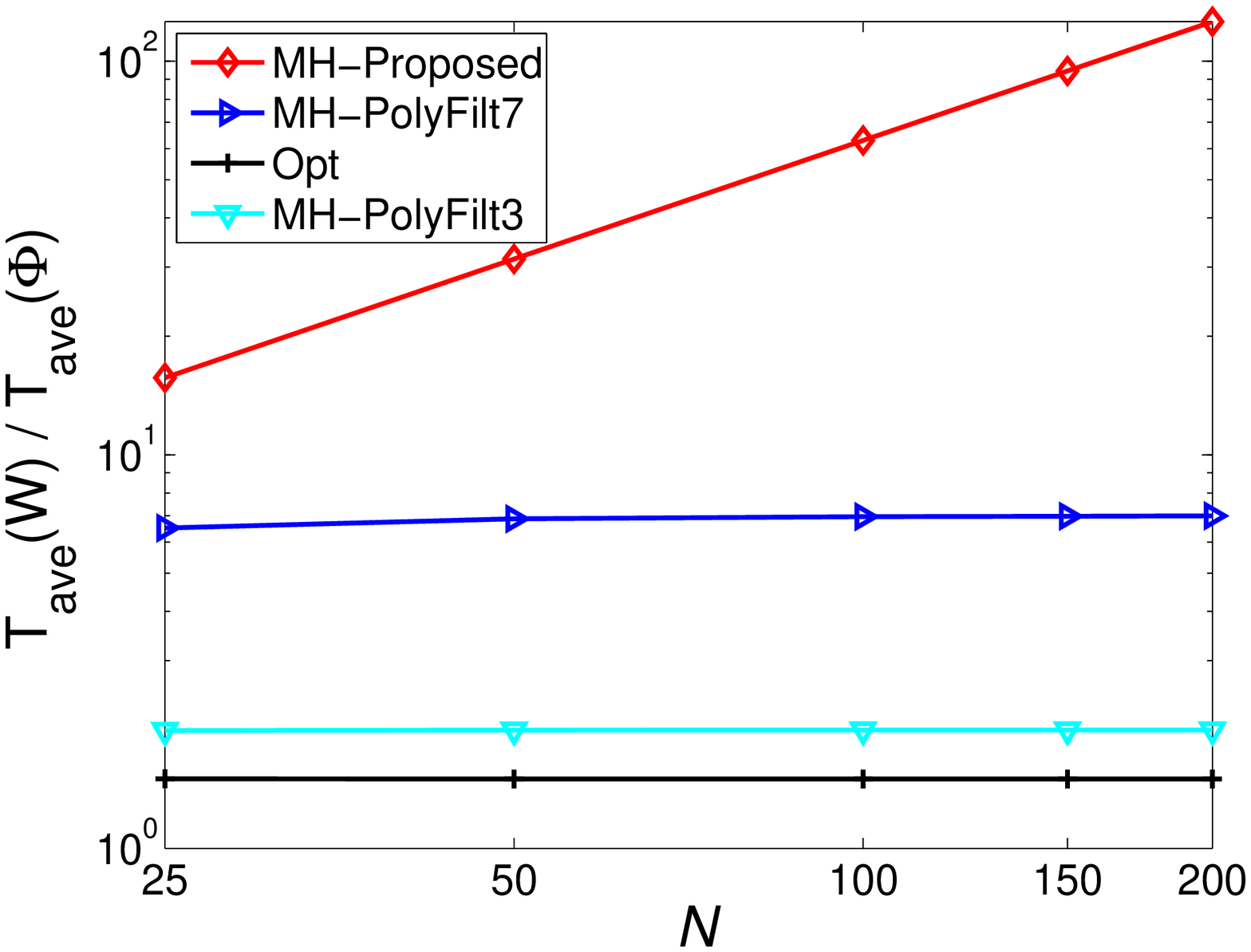}}
\caption{Averaging time characterization, chain topology. The
algorithms compared are: optimal weight matrix of Xiao and
Boyd~\cite{Xiao04} (Opt): $+$; polynomial filter with 3 taps
(MH-PolyFilt3): $\bigtriangledown$, and 7 taps (MH-PolyFilt7):
$\triangleright$; proposed method with oracle $\lambda_{2}(\bW)$ and
MH matrix (MH-Proposed): $\diamond$. (a) Averaging time as a
function of the network size. (b) Improvement due to the accelerated
consensus: ratio of the averaging time of the non-accelerated
algorithm to that of the associated accelerated algorithm.}
\label{fig:t_ave_circ} %% label for entire figure*
\end{figure*}

Figure~\ref{fig:mse_rg_circ_acc} compares the MSE curves for the
proposed algorithm with two versions of polynomial filtering
consensus~\cite{Kok09}, one using 3 taps and the other using 7 taps.
We see that in the RGG scenario, our algorithm outperforms
polynomial filtering with $3$ memory taps and converges at a rate
similar to that of the $7$-tap version of polynomial
filtering\footnote{Calculating optimal weights in the
polynomial filtering framework quickly becomes ill-conditioned with
increasing filter length, and we were not able to obtain stable
results for more than 7 taps on random geometric graph topologies.
Note that the original paper~\cite{Kok09} also focuses on filters of
length no more than $7$. We conjecture that this ill-conditioning stems from the fact that the optimal solution involves
pseudo-inversion of a Vandermonde matrix containing powers of the
original eigenvalues. Since, for random geometric graph topologies,
eigenvalues are not described by a regular function (e.g., the
cosine, as for the chain graph) there is a relatively high
probability (increasing with $N$) that the original weight matrix
contains two similar-valued eigenvalues which may result in
the Vandermonde matrix being ill-conditioned.}. Decentralized
calculation of topology-adapted polynomial filter weights also
remains an open problem.  We conclude that for random geometric
graphs, our algorithm has superior properties with respect to
polynomial filtering since it has better error performance for the
same computational complexity, and our approach is suitable for
completely distributed implementation. Moving our attention to the
chain topology only emphasizes these points, as our accelerated
algorithm significantly outperforms even 7-tap polynomial filtering.
Note that decentralized initialization of our algorithm also works
well in the chain graph scenario. However, to obtain this result we
have to increase the number of consensus iterations in the
eigenvalue estimation algorithm, $K$, from $2N$ to $N^2$. This
increase in the complexity of the distributed optimization of
accelerated consensus algorithm is due to the properties of the
power methods~\cite{Golub96} and related eigenvalue estimation
problems. The accuracy of the second largest eigenvalue computation
depends on the ratio $\lambda_{3}(\bW) / \lambda_{2}(\bW)$, and this
ratio increases much more rapidly for the chain topology as $N$
grows than it does for random geometric graphs.

To investigate the robustness and scalability properties of the
proposed algorithm, we next examine the averaging time,
$T_{\scriptsize \mbox{ave}}(\bPhi_3[\alpha^*])$, as defined in
\eqref{eqn:t_ave}, and the ratio $T_{\scriptsize
\mbox{ave}}(\bW)/T_{\scriptsize \mbox{ave}}(\bPhi_3[\alpha^*])$, for
random geometric graphs (Fig.~\ref{fig:t_ave_rgg}) and the chain
topology (Fig.~\ref{fig:t_ave_circ}). We establish through
simulation that the scaling behaviour of the ratio that can be
measured experimentally matches very well with the asymptotic result
established theoretically for the processing gain,
$\tau_{\asym}(\bW) / \tau_{\asym}(\bPhi_3[\alpha^*])$. We see from
Fig.~\ref{fig:t_ave_rgg} that in the random geometric graph setting,
the proposed algorithm always outperforms consensus with the optimal
weight matrix of Xiao and Boyd~\cite{Xiao04} and polynomial filter
with equal number of memory taps, and our approach scales comparably
to 7-tap polynomial filtering. On the other hand, in the chain graph
setting (Fig.~\ref{fig:t_ave_circ}) the proposed algorithm
outperforms all the competing algorithms.  Another interesting
observation from Fig.~\ref{fig:t_ave_circ} is that the gains of the
polynomial filter and optimal weight matrix remain almost constant
with varying network size while the gain obtained by the proposed
algorithm increases significantly with $N$.  This linear improvement
with $N$ matches well with the asymptotic behavior predicted by
Theorem~3.

\begin{figure*}[t]
\centering \subfigure[Random geometric graph]{
\label{fig:t_ave_circ_pap_acc:a} %% label for first subfigure
\includegraphics[width = 7.5cm]{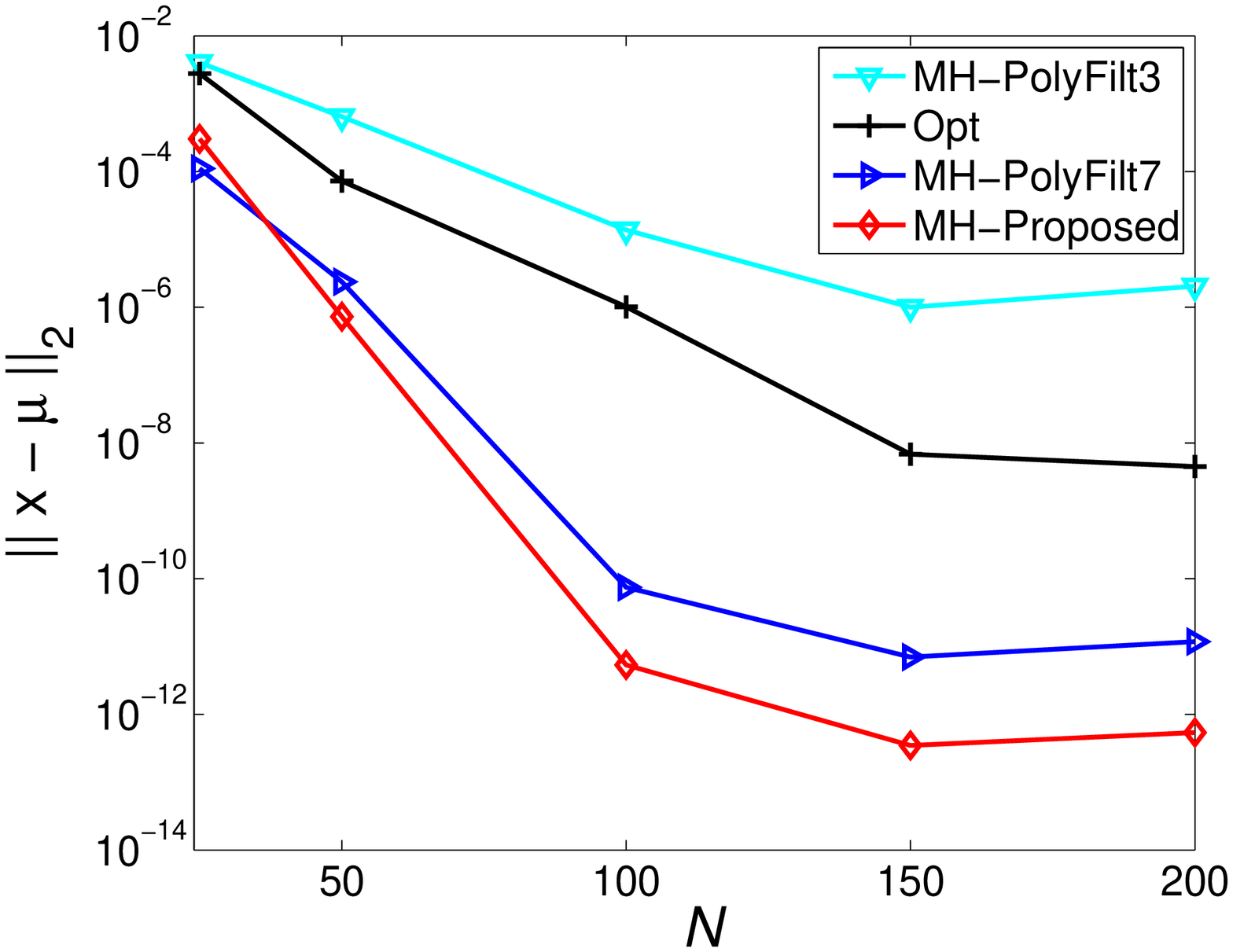}}
\hspace{1cm}
 \subfigure[Chain]{ % \hfill
\label{fig:t_ave_circ_pap_acc:b} %% label for second subfigure
\includegraphics[width = 7.5cm]{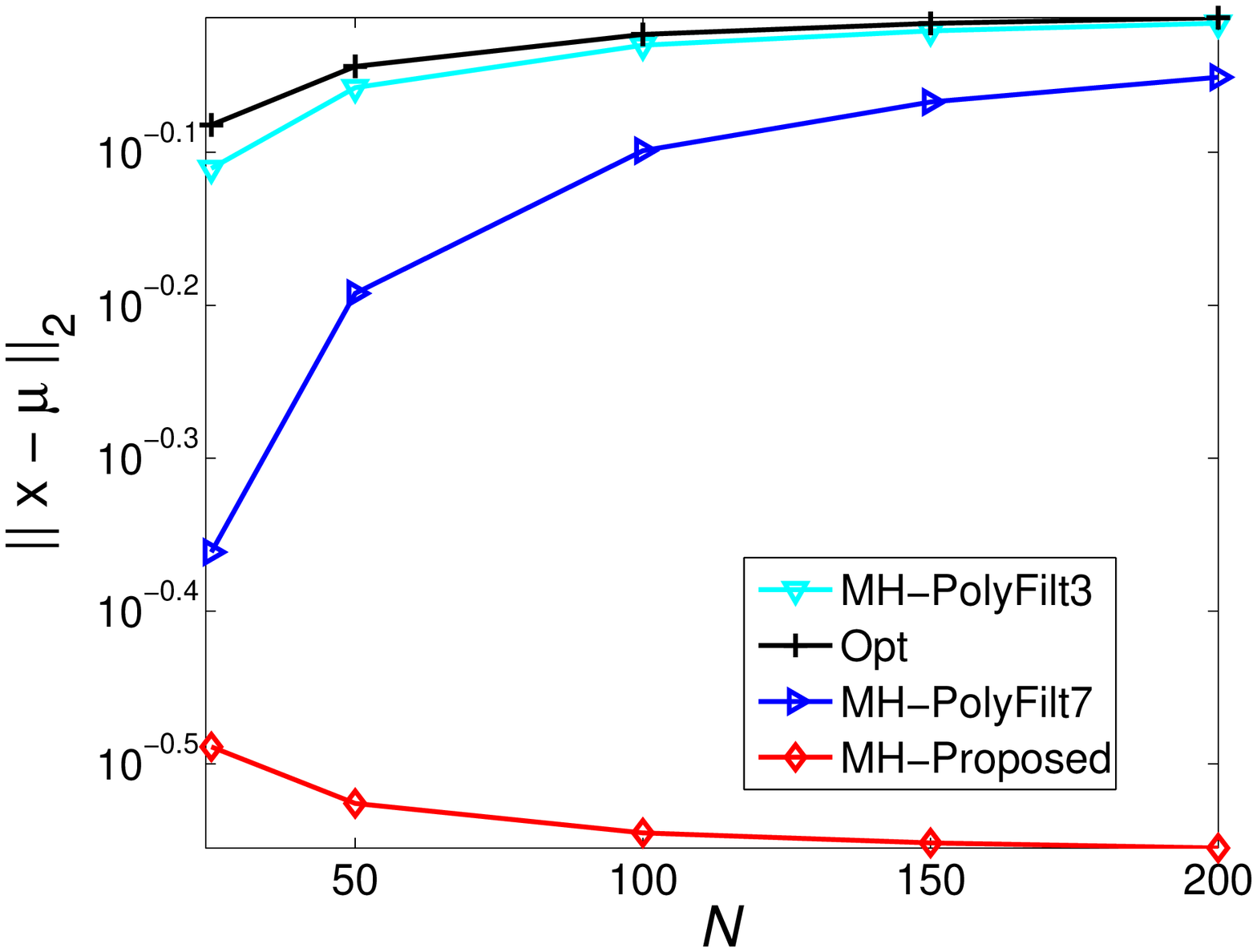}}
\caption{MSE at the point when finite time consensus of Sundaram and
Hadjicostis~\cite{Sundaram07} has enough information to calculate the exact
average at all nodes. The algorithms compared are: optimal weights (Opt): $+$; polynomial filter with 3 taps (MH-PolyFilt3): $\bigtriangledown$, and 7 taps (MH-PolyFilt7):
$\triangleright$; proposed method with oracle
$\lambda_{2}(\bW)$ and MH matrix (MH-Proposed): $\diamond$. (a) Random
geometric graph. (b) Chain topology.}
\label{fig:mse_hajicostis} %% label for entire figure*
\end{figure*}

Finally, we compare the proposed algorithm with the linear observer
approach of Sundaram and Hadjicostis~\cite{Sundaram07}, which works
by remembering all of the consensus values, $x_i(t)$, seen at a node
$i$ (unbounded memory).  After enough updates, each node is able to
perfectly recover the average by locally solving a set of linear
equations.  To compare the method of \cite{Sundaram07} with our
approach and the other asymptotic approaches described above, we
determine the topology-dependent number of iterations that the
linear-observer method must execute to have enough information to
exactly recover the average.  We then run each of the asymptotic
approaches for the same number of iterations and evaluate
performance based on the MSE they achieve.
Figure~\ref{fig:mse_hajicostis} depicts results for both random
geometric graph and chain topologies.  For random geometric graphs
of $N \ge 100$ nodes, we observe that the proposed algorithm
achieves an error of at most $10^{-12}$ (roughly machine precision),
by the time the linear observer approach has sufficient information
to compute the average.  For the chain topology the results are much
more favourable for the linear-observer approach. However, the
linear observer approach requires significant overhead to determine
the topology-dependent coefficients that define the linear system to
be solved at each node and does not scale well to large networks.

% Figure 5

\section{Proofs of Main Results and Discussion}
\label{section::Proofs}

\subsection{Limiting $\varepsilon$-convergence time}
\label{ssection::limiting_averaging_time}

To begin, we need to motivate choosing $\alpha$ to minimize the
spectral radius $\rho(\bPhi_3[\alpha] - \bJ)$ since, unlike in the
memoryless setting, it does not bound the step-wise rate of
convergence.  In fact, since $\bPhi_3[\alpha]$ is not symmetric,
$\bPhi_3[\alpha]^t$ does not even converge to $\bJ$ as $t
\rightarrow \infty$, as in the memoryless setting.  However, we will
show that: (i) for the proposed construction, $\bPhi_3[\alpha]^t$
does converge to a matrix $\bar{\bPhi}$; (ii) that the limiting
convergence time is governed by $\rho(\bPhi_3[\alpha] -
\bar{\bPhi})$; and (iii) that $\rho(\bPhi_3[\alpha] - \bar{\bPhi}) =
\rho(\bPhi_3[\alpha] - \bJ)$.

Before stating our first result we must introduce some notation. For
now, assume we are given a matrix $\bPhi \in \Re^{n \times n}$ with
$\bar{\bPhi} = \lim_{t \rightarrow \infty} \bPhi^t$.  We will
address conditions for existence of the limit below. For a given
initialization vector $\bx(0) \in \Re^n$, let $\tilde{\bx}(0) =
\bar{\bPhi} \bx(0)$, and define the set of non-trivial
initialization vectors $\mathcal{X}_{0,\bPhi} \triangleq \{ \bx(0)
\in \Re^n : \bx(0) \ne \tilde{\bx}(0) \}$.  Since we have not yet
established that $\tilde{\bx}(0) = \bar{\bx}(0) \triangleq \bJ
\bx(0)$, we keep the discussion general and use the following
definition of the convergence time:
\begin{align}
T_{\con}(\bPhi, \varepsilon) = \inf_{\tau \geq 0} \left\{ \tau : ||
\bx(t) - \tilde{\bx}(0) ||_2 \leq \varepsilon || \bx(0) -
\tilde{\bx}(0) ||_2 \quad \forall\ t \geq \tau , \quad \forall\
\bx(0) \in \mathcal{X}_{0,\bPhi} \right\}
\end{align}
We now prove a result relating the spectral radius and the
$\varepsilon$-convergence time for general non-symmetric averaging
matrices $\bPhi$, which we will then apply to our particular
construction, $\bPhi_3[\alpha]$.

\begin{theorem} \label{th:lim_time}
Let $\bPhi \in \mathbb{R}^{n \times n}$ be given, with limit $\lim_{t \rightarrow \infty} \bPhi^t = \bar \bPhi$, and assume that $\rho(\bPhi - \bar\bPhi) > 0$. Then
\begin{align} \label{eqn:avtimegeneral}
\lim\limits_{\varepsilon \rightarrow 0} \frac{T_{\con}(\bPhi,
\varepsilon)}{\log \varepsilon^{-1}} = \frac{1}{\log \rho(\bPhi -
\bar\bPhi)^{-1}}.
\end{align}
\end{theorem}
\begin{IEEEproof}
The limit $\lim_{t \rightarrow \infty} \bPhi^t =
\bar \bPhi$ exists if and only if (see~\cite{Mey77}) $\bPhi$
can be expressed in the form
\begin{align}
\bPhi = \T \left[\begin{array}{cc} \I_{\kappa} & \bZ \\
\bZ & \Z
\end{array}\right] \T^{-1}
\end{align}
where $\I_{\kappa}$ is the identity matrix of dimension $\kappa$,
$\Z$ is a matrix with $\rho(\Z) < 1$ and $\T$ is an invertible
matrix. It follows that in the limit we have~\cite{Joh08},
\begin{align}
\bar \bPhi = \lim_{t \rightarrow \infty} \bPhi^t = \T \left[\begin{array}{cc} \I_{\kappa} & \bZ \\
\bZ & \bZ
\end{array}\right] \T^{-1}.
\end{align}
By linear algebra, $\bPhi \bar \bPhi = \bar \bPhi \bPhi = \bar
\bPhi$ and $\bPhi^t \bar \bPhi = \bar \bPhi$. Using these facts it
is trivial to show $(\bPhi - \bar \bPhi)^t = \bPhi^t - \bar \bPhi$,
implying $(\bPhi - \bar \bPhi)^t(\bx(0) - \tilde{\bx}(0)) = \bx(t) -
\tilde{\bx}(0)$. Taking the norm of both sides we have
\begin{align}
|| \bx(t) - \tilde{\bx}(0) ||_2 = || (\bPhi - \bar\bPhi)^t (\bx(0) -
\tilde{\bx}(0)) ||_2,
\end{align}
and therefore
\begin{align} \label{eqn:T_conv_1}
T_{\con}(\bPhi, \varepsilon) = \inf_{\tau \geq 0} \left\{ \tau :
\frac{|| (\bPhi - \bar\bPhi)^t (\bx(0) - \tilde{\bx}(0)) ||_2}{||
\bx(0) - \tilde{\bx}(0) ||_2}  \leq \varepsilon  \quad \forall\ t
\geq \tau , \quad \forall\ \bx(0) \in \mathcal{X}_{0,\bPhi}
\right\}.
\end{align}
By the definition of $T_{\con}(\bPhi, \varepsilon)$ above we have:
\begin{equation}
\frac{|| (\bPhi - \bar\bPhi)^{T_{\con}(\bPhi, \varepsilon)} (\bx(0)
- \tilde{\bx}(0)) ||_2}{|| \bx(0) - \tilde{\bx}(0) ||_2} \leq
\varepsilon, \quad \forall \bx(0) \in \mathcal{X}_{0,\bPhi}.
\end{equation}
This implies:
\begin{equation}
\left[ \left(\sup\limits_{\bx(0) \in \mathcal{X}_{0,\bPhi}} \frac{||
(\bPhi - \bar\bPhi)^{T_{\con}(\bPhi, \varepsilon)} (\bx(0) -
\tilde{\bx}(0)) ||_2}{|| \bx(0) - \tilde{\bx}(0) ||_2}
\right)^{1/T_{\con}(\bPhi, \varepsilon)} \right]^{T_{\con}(\bPhi,
\varepsilon)} \leq \varepsilon, \label{eqn:toLog}
\end{equation}
and so, using the definition of the induced operator norm, which is
simply $|| \bPhi - \bar\bPhi ||_{2} = \sup_{\bx(0) \in
\mathcal{X}_{0,\bPhi}} || (\bPhi - \bar\bPhi) (\bx(0) -
\tilde{\bx}(0)) ||_2 / || \bx(0) - \tilde{\bx}(0)||_2$, after taking
the logarithm on both sides of \eqref{eqn:toLog}, we
have\footnote{Since we are interested in asymptotic behaviour of the
type $\varepsilon \rightarrow 0$, there is no loss of generality in
supposing that $\varepsilon$ is sufficiently small so that the
following holds: $\log \varepsilon < 0$, $\log || (\bPhi -
\bar\bPhi)^{T_{\con}(\bPhi, \varepsilon)} ||_{2}^{1/{T_{\con}(\bPhi,
\varepsilon)}} < 0$, and $\log || (\bPhi -
\bar\bPhi)^{T_{\con}(\bPhi, \varepsilon) - 1}
||_{2}^{1/{(T_{\con}(\bPhi, \varepsilon)- 1)}} < 0$}
\begin{align} \label{eqn:t_star_geq}
T_{\con}(\bPhi, \varepsilon) &\geq \frac{\log \varepsilon}{\log ||
(\bPhi - \bar\bPhi)^{T_{\con}(\bPhi, \varepsilon)}
||_{2}^{1/{T_{\con}(\bPhi, \varepsilon)}}}.
\end{align}
Since~\cite{hornJohnson} $\rho(\bPhi - \bar\bPhi) \leq || (\bPhi -
\bar\bPhi)^{t} ||_{2}^{1/t}$ for any $t \geq 0$, it follows that
\begin{align} \label{eqn:hjabsvjkhbvs}
T_{\con}(\bPhi, \varepsilon) &\geq \frac{\log \varepsilon}{\log
\rho(\bPhi - \bar\bPhi)}.
\end{align}
from which it is also clear that $T_{\con}(\bPhi, \varepsilon)
\rightarrow \infty$ as $\varepsilon \rightarrow 0$.

Now, by the definition of $T_{\con}(\bPhi, \varepsilon)$
in~\eqref{eqn:T_conv_1} we also have
\begin{equation}
\exists \bx(0) \in \mathcal{X}_{0,\bPhi}, \quad \frac{|| (\bPhi -
\bar\bPhi)^{T_{\con}(\bPhi, \varepsilon)-1} (\bx(0) -
\tilde{\bx}(0)) ||_2}{|| \bx(0) - \tilde{\bx}(0) ||_2} >
\varepsilon,
\end{equation}
implying, for the operator norm of $(\bPhi -
\bar\bPhi)^{T_{\con}(\bPhi, \varepsilon)-1}$:
\begin{align} \label{eqn:norm_grater_varepsilon}
||(\bPhi - \bar\bPhi)^{T_{\con}(\bPhi, \varepsilon)-1}||_{2}
> \varepsilon.
\end{align}
From \eqref{eqn:norm_grater_varepsilon} and \eqref{eqn:toLog} it
follows $||(\bPhi - \bar\bPhi)^{T_{\con}(\bPhi, \varepsilon)}||_{2}
\leq \varepsilon < ||(\bPhi - \bar\bPhi)^{T_{\con}(\bPhi,
\varepsilon)-1}||_{2}$ and thus we can always pick $\beta \in [0,1)$
such that the following holds:
\begin{align}
\beta ||(\bPhi - \bar\bPhi)^{T_{\con}(\bPhi, \varepsilon)-1}||_{2} +
(1-\beta) ||(\bPhi - \bar\bPhi)^{T_{\con}(\bPhi, \varepsilon)}||_{2}
= \varepsilon.
\end{align}
Using the notation $C_{T_{\con}} = ||(\bPhi -
\bar\bPhi)^{T_{\con}(\bPhi, \varepsilon)}||_{2} / ||(\bPhi -
\bar\bPhi)^{T_{\con}(\bPhi, \varepsilon)-1}||_{2}$ for the bounded
number $C_{T_{\con}}$ we conclude
\begin{align} \label{eqn:kjzvkj}
||(\bPhi - \bar\bPhi)^{T_{\con}(\bPhi, \varepsilon)-1}||_{2} ( \beta
+ (1-\beta) C_{T_{\con}} )  = \varepsilon.
\end{align}
The boundedness of $C_{T_{\con}}$ follows from the
sub-multiplicativity of the operator norm, $||(\bPhi -
\bar\bPhi)^{T_{\con}(\bPhi, \varepsilon)}||_{2} \leq \|\bPhi -
\bar\bPhi \|_2 ||(\bPhi - \bar\bPhi)^{T_{\con}(\bPhi,
\varepsilon)-1}||_{2}$ yielding $0 \leq C_{T_{\con}} \leq \|\bPhi -
\bar\bPhi \|_2$.

Using the technique used to switch from \eqref{eqn:toLog} to
\eqref{eqn:t_star_geq} we obtain from \eqref{eqn:kjzvkj}:
\begin{align}
(T_{\con}(\bPhi, \varepsilon)-1) \log ||(\bPhi -
\bar\bPhi)^{T_{\con}(\bPhi,
\varepsilon)-1}||_{2}^{1/(T_{\con}(\bPhi, \varepsilon)-1)} = \log
\varepsilon - \log(\beta  + (1-\beta) C_{T_{\con}} ).
\end{align}
Dividing through by $\log\varepsilon^{-1} \log ||(\bPhi -
\bar\bPhi)^{T_{\con}(\bPhi,
\varepsilon)-1}||_{2}^{1/(T_{\con}(\bPhi, \varepsilon)-1)}$, and
taking the limit as $\varepsilon \rightarrow 0$ we have
\begin{align}
\lim\limits_{\varepsilon \rightarrow 0} \frac{T_{\con}(\bPhi,
\varepsilon)}{\log\varepsilon^{-1}} &= \lim\limits_{\varepsilon
\rightarrow 0} \frac{-1}{\log ||(\bPhi - \bar\bPhi)^{T_{\con}(\bPhi,
\varepsilon)-1}||_{2}^{1/(T_{\con}(\bPhi, \varepsilon)-1)}} \nonumber\\
&- \lim\limits_{\varepsilon \rightarrow 0} \frac{\log(\beta  +
(1-\beta) C_{T_{\con}} )}{\log ||(\bPhi -
\bar\bPhi)^{T_{\con}(\bPhi,
\varepsilon)-1}||_{2}^{1/(T_{\con}(\bPhi,
\varepsilon)-1)}\log\varepsilon^{-1}} + \lim\limits_{\varepsilon
\rightarrow 0} \frac{1}{\log\varepsilon^{-1}}.
\end{align}
Moving the limits on the right under the logs and using the fact
that $T_{\con}(\bPhi, \varepsilon) \rightarrow \infty$ as
$\varepsilon \rightarrow 0$, we may employ Gelfand's
formula~\cite{hornJohnson}, $\lim_{t \rightarrow \infty} || (\bPhi -
\bar\bPhi)^t ||^{1/t} = \rho(\bPhi - \bar\bPhi)$:
\begin{align}
\lim\limits_{\varepsilon \rightarrow 0} \frac{T_{\con}(\bPhi,
\varepsilon)}{\log\varepsilon^{-1}} &= \frac{-1}{\log
\lim\limits_{\varepsilon \rightarrow 0} ||(\bPhi -
\bar\bPhi)^{T_{\con}(\bPhi,
\varepsilon)-1}||_{2}^{1/(T_{\con}(\bPhi, \varepsilon)-1)}} -
\lim\limits_{\varepsilon \rightarrow 0} \frac{\log(\beta  +
(1-\beta) C_{T_{\con}} )}{\log \lim\limits_{\varepsilon \rightarrow
0} ||(\bPhi - \bar\bPhi)^{T_{\con}(\bPhi,
\varepsilon)-1}||_{2}^{1/(T_{\con}(\bPhi,
\varepsilon)-1)}\log\varepsilon^{-1}} \nonumber \\
&= \frac{1}{\log \rho(\bPhi - \bar\bPhi)^{-1}} +
\lim\limits_{\varepsilon \rightarrow 0} \frac{\log(\beta  +
(1-\beta) C_{T_{\con}} )}{\log \rho(\bPhi - \bar\bPhi)^{-1}
\log\varepsilon^{-1}}
\end{align}
Using $C_{T_{\con}} \leq \|\bPhi - \bar\bPhi\|$, $\log(\beta  +
(1-\beta) C_{T_{\con}} ) \leq |\log(\beta  + (1-\beta) \|\bPhi -
\bar\bPhi\| )|$, and taking into account the fact that $0 \leq
|\log(\beta + (1-\beta) \|\bPhi - \bar\bPhi\| )| \leq |\log\|\bPhi -
\bar\bPhi\| |, \  \forall \beta \in [0,1],$ we have
\begin{align}
\lim\limits_{\varepsilon \rightarrow 0} \frac{T_{\con}(\bPhi,
\varepsilon)}{\log\varepsilon^{-1}} &\leq \frac{1}{\log \rho(\bPhi -
\bar\bPhi)^{-1}} + \lim\limits_{\varepsilon \rightarrow 0}
\frac{|\log(\|\bPhi - \bar\bPhi\| )|}{\log \rho(\bPhi -
\bar\bPhi)^{-1}
\log\varepsilon^{-1}} \nonumber\\
&= \frac{1}{\log \rho(\bPhi - \bar\bPhi)^{-1}}
\end{align}
Combining the last inequality with~\eqref{eqn:hjabsvjkhbvs}
completes the proof.
\end{IEEEproof}

In order to apply the above result, we must establish that
$\bPhi_3[\alpha]$  satisfies the conditions of
Theorem~\ref{th:lim_time}.  In doing so, we will also show that (i)
for $\bPhi = \bPhi_3[\alpha]$ and $\bX(0)$ defined in~\eqref{eqn:X},
the limit $\bar\bPhi \bX(0) = \bJ \bX(0)$, so our approach indeed
converges to the average consensus, and (ii) that the limiting
convergence time is characterized by a function of
$\rho(\bPhi_3[\alpha] - \bJ)$, which motivates choosing $\alpha$ to
optimize this expression. (Recall, in this setting $\bJ$ is the $2N
\times 2N$ matrix with all entries equal to $1/2N$.)  Note that the
condition on $\alpha$ is necessary for $\bPhi_3[\alpha]^t$ to have a
limit as $t\rightarrow \infty$, as will be established in
Section~\ref{ssection::OptParamProof}.

\begin{proposition}\label{prop:lim_time}
  Let $\bPhi_3[\alpha]$ be defined as in (\ref{eqn:operator_phi_M}), assume that the assumptions of Theorem~\ref{lem:optimization} hold, and $\alpha \in [0,-\theta_{1}^{-1})$.
  Then:
  \begin{description}
\item[(a)] $\bar\bPhi_3[\alpha] = \lim_{t \rightarrow \infty} \bPhi_3[\alpha]^t$ exists,
  with $\bar\bPhi_3[\alpha] \bX(0) = \bJ \bX(0)$ for all $\bX(0)$ defined in~\eqref{eqn:X},
\item[(b)] $\rho(\bPhi_3[\alpha] -
  \bar\bPhi_3[\alpha]) > 0$, and \item[(c)]
%\begin{align} \label{eqn:avtimespecific}
$\lim\limits_{\varepsilon \rightarrow 0}
\frac{T_{\con}(\bPhi_3[\alpha], \varepsilon)}{\log \varepsilon^{-1}}
= \frac{1}{\log \rho(\bPhi_3[\alpha] - \J)^{-1}}.$
%\end{align}
\end{description}
\end{proposition}

\begin{IEEEproof}  \textit{Proof of part (a).}
In Theorem~1 in~\cite{Joh08}, Johansson and  Johansson show that the
necessary and sufficient conditions for the consensus algorithm of
the form $\bPhi_{3}[\alpha]$ to converge to the average are (JJ1)
$\bPhi_{3}[\alpha] \1 = \1$; (JJ2)~$\bg^T \bPhi_{3}[\alpha] = \bg^T$
for vector $\bg^T = [\beta_1 \1^T \beta_2 \1^T]$ with weights
satisfying $\beta_1 + \beta_2 = 1$; and
(JJ3)~$\rho(\bPhi_{3}[\alpha] - \frac{1}{N} \1 \bg^T) < 1$.  If
these conditions hold then we also have $\bar\bPhi_3[\alpha] =
\frac{1}{N} \1 \bg^T$~\cite{Joh08} implying $\tilde{\bX}(0) =
\bar{\bX}(0)$.  Condition (JJ1) is easily verified after
straightforward algebraic manipulations using the definition of
$\bPhi_3[\alpha]$ in \eqref{eqn:operator_phi_M}, the assumption that
$\theta_1 + \theta_2 + \theta_3 = 1$, and recalling that $\bW$
satisfies $\bW \1 = \1$ by design.  To address condition (JJ2), we
set $\beta_1 = 1/(1 + \alpha \theta_1)$ and $\beta_2 =
\alpha\theta_1 / (1 + \alpha \theta_1)$.  Clearly, $\beta_1 +
\beta_2 = 1$, and it is also easy to verify condition (JJ2) by
plugging these values into the definition of $\bg$, and using the
same properties of $\bPhi_3[\alpha]$, the $\theta_i$'s, and $\bW$ as
above.

In order to verify that condition (JJ3) holds, we will show here
that $\rho(\bPhi_3[\alpha] - \frac{1}{N}\1\bg^T) =
\rho(\bPhi_3[\alpha] - \bJ)$.  In
Section~\ref{ssection::OptParamProof} we show that
$\rho(\bPhi_3[\alpha] - \bJ) < 1$ if $\alpha \in
[0,-\theta_1^{-1})$, and thus condition (JJ3) is also satisfied
under the assumptions of the proposition.  To show that
$\rho(\bPhi_3[\alpha] - \frac{1}{N}\1 \bg^T) = \rho(\bPhi_3[\alpha]
- \bJ)$, we prove a stronger result, namely that
$\bPhi_3[\alpha]-\frac{1}{N}\1\bg^T$ and $\bPhi_3[\alpha] - \bJ$
have the same eigenspectra.  Consider the eigenvector $\bv_i$ of
$\bPhi_{3}[\alpha]$ with corresponding eigenvalue
$\lambda_i(\bPhi_3[\alpha])$.  This pair solves the eigenvalue
problem, $\bPhi_3[\alpha]\bv_i = \lambda_i(\bPhi_3[\alpha]) \bv_i$.
Equivalently, expanding the definition of $\bPhi_3[\alpha]$, we have
\begin{align} \label{eqn:lenear_problem}
\left[\begin{array}{cc} \bW_{3}[\alpha] & \alpha\theta_{1}\bI \\
\bI & \bZ
\end{array}\right] \bv_i = \lambda_i(\bPhi_3[\alpha]) \left[\begin{array}{cc} \bI & \bZ \\
\bZ & \bI
\end{array}\right] \bv_i.
\end{align}
We observe that (\ref{eqn:lenear_problem}) fits  a modification of
the first companion form of the linearization of a Quadratic
Eigenvalue Problem (QEP) (see Section 3.4 in~\cite{Tisseur01}). The
QEP has general form $(\lambda^2\bM + \lambda\bC + \bK) \bu = \bZ$,
where $\bu$ is the eigenvector associated with this QEP.  The
linearization of interest to us has the form:
\begin{align} \label{eqn:lenear_problem2}
\left[\begin{array}{cc} -\bC & -\bK \\
\bI & \bZ
\end{array}\right]  \left[\begin{array}{c} \lambda \u \\
\u
\end{array}\right] - \lambda \left[\begin{array}{cc} \bM & \bZ \\
\bZ & \bI
\end{array}\right]  \left[\begin{array}{c} \lambda \u \\
\u
\end{array}\right] = \bZ.
\end{align}
The correspondence is clear if we make the associations: $\bM =
\bI$,  $\bC = -\bW_3[\alpha]$ and $\bK = -\alpha\theta_{1}\bI$,
$\lambda = \lambda_i(\bPhi_3[\alpha])$ and $\bv_i =
[\lambda_i(\bPhi_3[\alpha]) \u^T \u^T]^T$. Eigenvectors $\bv_i$ that
solve (\ref{eqn:lenear_problem}) thus have special structure and are
related to $\bu_i$, the solution to the QEP,
\begin{align} \label{eqn:QEP0}
(\lambda_i(\bPhi_3[\alpha])^2\bI -
\lambda_i(\bPhi_3[\alpha]) \bW_{3}[\alpha] -\alpha\theta_{1}\bI) \bu_i
= \bZ.
\end{align}
Because the  first and third terms above are scaled identity
matrices and the definition of $\bW_3[\alpha]$
(see~\eqref{eqn:Wmalpha}) also involves scaled identity matrices, we
can simplify this last equation to find that any solution $\bu_i$
must also be an eigenvector of $\bW$.

We have seen  above, when verifying condition (JJ1), that $\1$ is an
eigenvector of $\bPhi_3[\alpha]$ with corresponding eigenvalue
$\lambda_i(\bPhi_3[\alpha]) = 1$. Observe that, from the definition
of $\bg$ and because $\beta_1 + \beta_2 = 1$, we have
$(\frac{1}{N}\1 \bg^T) \1 = \1$. Thus, $(\bPhi_3[\alpha] -
\frac{1}{N}\1 \bg^T)\1 = \bZ$.  Similarly, recalling that $\bJ =
\frac{1}{2N}\1\1^T$, we have $\bJ \1 = \1$, and thus
$(\bPhi_3[\alpha] - \bJ)\1 = \bZ$.  By design, $\bW$ is a doubly
stochastic matrix, and all eigenvectors $\bu$ of $\bW$ with $\bu \ne
\1$ are orthogonal to $\1$.  It follows that $(\frac{1}{N}\1
\bg^T)\bv_i = \bZ$ for corresponding eigenvectors $\bv_i =
[\lambda_i(\bPhi_3[\alpha]) \bu^T \bu^T]^T$ of $\bPhi_3[\alpha]$,
and thus $(\bPhi_3[\alpha] - \frac{1}{N}\1\bg^T)\bv_i =
\bPhi_3[\alpha] \bv_i = \lambda_i(\bPhi_3[\alpha]) \bv_i$.
Similarly, $\bJ \bv_i = 0$ if $\bv_i \ne \1$, and $(\bPhi_3[\alpha]
- \bJ) \bv_i = \lambda_i(\bPhi_3[\alpha]) \bv_i$.  Therefore, we
conclude that the matrices $(\bPhi_3[\alpha] - \bar\bPhi_3[\alpha])$
and $(\bPhi_3[\alpha] - \bJ)$ have identical eigenspectra, and thus
$\rho(\bPhi_3[\alpha] - \frac{1}{N}\1\bg^T) = \rho(\bPhi_3[\alpha] -
\bJ)$.

In Section~\ref{ssection::OptParamProof} we  show that
$\rho(\bPhi_3[\alpha] - \bJ) < 1$ if $\alpha \in
[0,-\theta_1^{-1})$, and thus the assumptions of the proposition,
taken together with the analysis just conducted, verify that
condition (JJ3) is also satisfied.  Therefore, the limit
$\lim_{t\rightarrow\infty} \bPhi_3[\alpha]^t = \bar\bPhi_3[\alpha] =
\frac{1}{N}\1\bg^T$ exists, and $\bar\bPhi_3[\alpha] \bX(0) = \bJ
\bX(0)$ for all $\bX(0)$ defined in~\eqref{eqn:X}.

\textit{Proofs of parts (b) and (c).} In the proof of
Lemma~\ref{lem:remJ1} (see Section~\ref{ssection::OptParamProof}),
it is shown that $\rho(\bPhi_{3}[\alpha] - \J]) \geq -\alpha
\theta_{1}$.  Thus, if $\alpha > 0$ and $\theta_1 < 0$, then part
(b) holds.  The assumptions $\theta_1 + \theta_2 + \theta_3 = 1$,
$\theta_3 \ge 1$, and $\theta_2 \ge 0$ imply that $\theta_1 \le 0$,
and by assumption, $\alpha \ge 0$.  If $\alpha = 0$ or $\theta_1 =
0$, then the proposed predictive consensus scheme reduces to
memoryless consensus with weight matrix $\bW$ (and the statement
follows directly from the results of \cite{Xiao04,Boyd06}).  Thus,
part (b) of the proposition follows from the assumptions and the
analysis in Lemma~\ref{lem:remJ1} below.  By proving parts (a) and
(b), we have verified the assumptions of Theorem~\ref{th:lim_time}
above.  Applying the result of this Theorem, together with the
equivalence of $\rho(\bPhi_3[\alpha] - \frac{1}{N}\1\bg^T)$ and
$\rho(\bPhi_3[\alpha] - \bJ)$, gives the claim in part (c), thereby
completing the proof.
\end{IEEEproof}

\subsection{Proof of Theorem~\ref{lem:optimization}: Optimal Mixing Parameter}
\label{ssection::OptParamProof}

In order to minimize the spectral radius of $\bPhi_{3}[\alpha]$ we
need to know its eigenvalues. These can be calculated by solving the
eigenvalue problem (\ref{eqn:lenear_problem}).  We can multiply (\ref{eqn:QEP0}) by $\bu^T_{i}$ on the left
to obtain a quadratic equation that links the individual eigenvalues
$\lambda_i(\bPhi_{3}[\alpha])$ and $\lambda_i(\bW_{3}[\alpha])$:
\begin{align} \label{eqn:QEP}
\bu_{i}^T(\lambda_i(\bPhi_{3}[\alpha])^2\bI -
\lambda_i(\bPhi_{3}[\alpha])\bW_{3}[\alpha]
-\alpha\theta_{1}\bI) \bu_{i} &= 0 \nonumber \\
\lambda_i(\bPhi_{3}[\alpha])^2 - \lambda_i(\bW_{3}[\alpha])
\lambda_i(\bPhi_{3}[\alpha]) -\alpha\theta_{1} &= 0.
\end{align}
Recall $\bPhi_3[\alpha]$ is a $2N \times 2N$ matrix, and so
$\bPhi_3[\alpha]$ has, in general, $2N$ eigenvalues -- twice as many
as $\bW_3[\alpha]$.  These eigenvalues are the solutions of the
quadratic (\ref{eqn:QEP}), and are given by
\begin{align} \label{eqn:roots_lam_plus}
\lambda_i^*(\bPhi_{3}[\alpha]) &= \frac{1}{2}
\left(\lambda_i(\bW_{3}[\alpha]) +
\sqrt{\lambda_i(\bW_{3}[\alpha])^2 + 4 \alpha \theta_{1}}\right) \nonumber \\
\lambda_i^{**}(\bPhi_{3}[\alpha]) &= \frac{1}{2}
\left(\lambda_i(\bW_{3}[\alpha]) -
\sqrt{\lambda_i(\bW_{3}[\alpha])^2 + 4 \alpha \theta_{1}}\right).
\end{align}
With these expressions for the eigenvalues of $\bPhi_{3}[\alpha]$, we
are in a position to formulate the problem of minimizing
the spectral radius of the matrix $(\bPhi_{3}[\alpha] - \bJ)$,
%\begin{align} \label{eqn:opt_prob_start}
$\alpha^{\star} = \arg \min\limits_{\alpha} \rho(\bPhi_{3}[\alpha] -
\bJ)$.
%\end{align}
It can be shown that this problem is equivalent to
\begin{align} \label{eqn:opt_prob_start_1}
\alpha^{\star} = \arg \min\limits_{\alpha \geq 0}
\rho(\bPhi_{3}[\alpha] - \bJ)
\end{align}
The simplest way to demonstrate this is to show that
$\rho(\bPhi_{3}[\alpha] - \bJ) \geq \rho(\bPhi_{3}[0] - \bJ)$ for
any $\alpha < 0$. Indeed, by the definition of the spectral radius
we have that $\rho(\bPhi_{3}[\alpha] - \bJ) \geq
\lambda_2^*(\bPhi_{3}[\alpha])$ and $\rho(\bPhi_{3}[0] - \bJ) =
\lambda_2(\bW)$. The latter is clear if we plug $\alpha=0$ into
\eqref{eqn:roots_lam_plus}. Hence it is enough to demonstrate
$\lambda_2^*(\bPhi_{3}[\alpha]) \geq \lambda_2(\bW)$. Consider the
inequality $\lambda_2^*(\bPhi_{3}[\alpha]) - \lambda_2(\bW) \geq 0.$
Replacing $\lambda_2^*(\bPhi_3[\alpha])$ with its definition
according to \eqref{eqn:roots_lam_plus}, rearranging terms and
squaring both sides gives $\alpha \theta_1 \ge \lambda_2(\bW)^2 -
\lambda_2(\bW) \lambda_2(\bW_3[\alpha])$.  From the definition of
$\bW_3[\alpha]$ in \eqref{eqn:Wmalpha}, it follows that
$\lambda_{2}(\bW_3[\alpha]) = (1 - \alpha + \alpha \theta_3)
\lambda_{2}(\bW) + \alpha \theta_2$. Using this relation leads to
the expression $\alpha(\theta_1 + (\theta_3 - 1)\lambda_2(\bW)^2 +
\theta_2 \lambda_2(\bW)) \ge 0$. Under our assumptions, we have
$\theta_{3}-1 \geq 0$, $\theta_{2} \geq 0$ and $\theta_{1} \leq 0$.
Thus $\theta_{1} + (\theta_{3}-1)\lambda_{2}(\bW)^2 +
\theta_{2}\lambda_{2}(\bW) \leq \theta_{1} + \theta_{3}-1 +
\theta_{2} = 0$ since $\lambda_{2}(\bW) < 1$. This implies that if
$\alpha < 0$, the last inequality holds leading to
$\lambda_2^*(\bPhi_{3}[\alpha]) \geq \lambda_2(\bW)$. Thus for any
$\alpha < 0$ the spectral radius $\rho(\bPhi_{3}[\alpha] - \bJ)$
cannot decrease, and so we may focus on optimizing over $\alpha \ge
0$.
% showing that (\ref{eqn:opt_prob_start}) and (\ref{eqn:opt_prob_start_1}) are indeed equivalent.

Now, the proof of Theorem~\ref{lem:optimization} boils down to
examining how varying $\alpha$ affects the eigenvalues of
$\bPhi_3[\alpha]$ on a case-by-case basis.  We first show that the
first eigenvalues, $\lambda^*_1(\bPhi_3[\alpha])$ and
$\lambda^{**}_1(\bPhi_3[\alpha])$, are smaller than all the others.
Then, we demonstrate that the second eigenvalues,
$\lambda^*_2(\bPhi_3[\alpha])$ and
$\lambda^{**}_2(\bPhi_3[\alpha])$, dominate all other pairs,
$\lambda^*_j(\bPhi_3[\alpha])$ and
$\lambda^{**}_j(\bPhi_3[\alpha])$, for $j > 2$, allowing us to focus
on the second eigenvalues, from which the proof follows.  Along the
way, we establish conditions on $\alpha$ which guarantee stability
of the proposed two-tap predictive consensus methodology.

To begin, we reformulate the optimization problem in terms of the
eigenvalues of $\bPhi_{3}[\alpha]$. We first consider
$\lambda_{1}^{*}(\bPhi_{3}[\alpha])$ and
$\lambda_{1}^{**}(\bPhi_{3}[\alpha])$. Substituting
$\lambda_{1}(\bW_{3}[\alpha]) = (1-\alpha+\alpha\theta_{3}) +
\alpha\theta_{2}$ we obtain the relationship
$\sqrt{\lambda^2_{1}(\bW_{3}[\alpha]) + 4 \alpha \theta_{1}} = |1 +
\alpha\theta_{1}|$ and using the condition $\theta_{1} \leq 0$, we
conclude that
\begin{align}
\lambda_{1}^{*}(\bPhi_{3}[\alpha]),
\lambda_{1}^{**}(\bPhi_{3}[\alpha]) &= \left\{\begin{array}{cc} 1,
-\alpha\theta_{1}& \textrm{if \ } 1 +
\alpha\theta_{1} \geq 0 \Rightarrow \alpha \leq -\theta_{1}^{-1}\\
-\alpha\theta_{1}, 1 & \textrm{if \ } 1 + \alpha\theta_{1} < 0
\Rightarrow \alpha > -\theta_{1}^{-1}.
\end{array}\right.
\end{align}
We note that $\alpha > -\theta_{1}^{-1}$ implies
$|\lambda_{1}^{**}(\bPhi_{3}[\alpha])| > 1$, leading to divergence
of the linear recursion involving $\bPhi_3[\alpha]$, and thus
conclude that the potential solution is restricted to the range
$\alpha \leq -\theta_{1}^{-1}$.  Focusing on this setting, we write
$\lambda_{1}^{*}(\bPhi_{3}[\alpha]) = 1$ and
$\lambda_{1}^{**}(\bPhi_{3}[\alpha]) = -\alpha\theta_{1}$.  We can
now reformulate the problem (\ref{eqn:opt_prob_start_1}) in terms of
the eigenvalues of $\bPhi_{3}[\alpha]$:
\begin{align} \label{eqn:opt_prob_eig}
\alpha^{\star} = \arg \min\limits_{\alpha \geq 0}
\max\limits_{i=1,2,\ldots N} \mathcal{J}_i[\alpha, \lambda_{i}(\bW)]
\end{align}
where
\begin{equation}
\mathcal{J}_i[\alpha, \lambda_{i}(\bW)] =
\begin{cases}
|\lambda_{1}^{**}(\bPhi_{3}[\alpha])|, & i=1\\
\max
(|\lambda_{i}^{*}(\bPhi_{3}[\alpha])|,|\lambda_{i}^{**}(\bPhi_{3}[\alpha])|)
& i>1.\end{cases}
\end{equation}

We now state a lemma that characterizes the functions
$\mathcal{J}_i[\alpha, \lambda_{i}(\bW)]$.

\begin{lemma} \label{lem:remJ1}
Under the assumptions of Theorem~\ref{lem:optimization},
\begin{align} \label{eqn:partial_cost}
\mathcal{J}_i[\alpha, \lambda_{i}(\bW)] &= \left\{\begin{array}{cc}
\alpha^{1/2} (-\theta_{1})^{1/2} & \textrm{if \ } \alpha \in [\alpha^{*}_i, \theta_{1}^{-1}]\\
\frac{1}{2} \left( \left|\lambda_{i}(\bW_{3}[\alpha])\right| +
\sqrt{\lambda_{i}(\bW_{3}[\alpha])^2 + 4 \alpha \theta_{1}} \right)
& \textrm{if \ } \alpha \in [0, \alpha^{*}_i)
\end{array}\right.
\end{align}
where
\begin{equation}
\alpha^{*}_i = \frac{-((\theta_{3}-1) \lambda_{i}(\bW)^2 +
\theta_{2} \lambda_{i}(\bW) + 2\theta_{1}) - 2\sqrt{\theta_{1}^2 +
\theta_{1}\lambda_{i}(\bW)\left(\theta_{2}+(\theta_{3}-1)\lambda_{i}(\bW)
\right)}}{\left(\theta_{2} + (\theta_{3}-1)\lambda_{i}(\bW)\right)^2
} \label{eqn:alpha_roots_1}
\end{equation}
Over the range $\alpha \in [0,-\theta_{1}^{-1}]$,
$\mathcal{J}_i[\alpha, \lambda_{i}(\bW)]\geq \mathcal{J}_1[\alpha,
\lambda_{1}(\bW)]$ for $i=2,3,\dots,N$.
\end{lemma}

\begin{proof}
For $i=2,3,\ldots N$, the eigenvalues
$\lambda_{i}^{*}(\bPhi_{3}[\alpha])$ and
$\lambda_{i}^{**}(\bPhi_{3}[\alpha])$ can admit two distinct forms; when the expression under the square root in
(\ref{eqn:roots_lam_plus}) is less then zero,
the respective eigenvalues are complex, and when this expression is
positive, the eigenvalues are real. In the region where the eigenvalues are complex,
\begin{align} \label{eqn:lam_plus_comp}
\max ( |\lambda_{i}^{*}(\bPhi_{3}[\alpha])|,
|\lambda_{i}^{**}(\bPhi_{3}[\alpha])| ) &=\frac{1}{2} \left[
\lambda_{i}(\bW_{3}[\alpha])^2 + \imath^2
\left(\sqrt{\lambda_{i}(\bW_{3}[\alpha])^2 + 4 \alpha
\theta_{1}}\right)^2
\right]^{1/2} \nonumber \\
&= \alpha^{1/2} (-\theta_{1})^{1/2}.
\end{align}
We note that (\ref{eqn:lam_plus_comp}) is a strictly increasing
function of $\alpha$. Recalling that $\lambda_{i}(\bW_{3}[\alpha]) =
(1+\alpha(\theta_{3}-1))\lambda_{i}(\bW) + \alpha\theta_{2}$ and
solving the quadratic $\lambda_{i}(\bW_{3}[\alpha])^2 + 4 \alpha
\theta_{1} = 0$, we can identify region, $[\alpha^*_i,
\alpha^{**}_i]$, where the eigenvalues are complex. The upper
boundary of this region is
\begin{equation}
\alpha^{**}_i = \frac{-((\theta_{3}-1) \lambda_{i}(\bW)^2 +
\theta_{2} \lambda_{i}(\bW) + 2\theta_{1}) +
2\sqrt{\theta_{1}^2+\theta_{1}\lambda_{i}(\bW)\left(\theta_{2}+(\theta_{3}-1)\lambda_{i}(\bW)\right)
}}{\left(\theta_{2} + (\theta_{3}-1)\lambda_{i}(\bW)\right)^2
} \label{eqn:alpha_roots_2}
\end{equation}
Relatively straightforward algebraic manipulation of
(\ref{eqn:alpha_roots_1}) and (\ref{eqn:alpha_roots_2}) leads to the
following conclusion:
if $\lambda_{i}(\bW) \in
[-1,1]$, $\theta_2\geq 0$ and $\theta_3\geq 1$, then $0 \leq
\alpha^{*}_i \leq -\theta_1^{-1} \leq \alpha^{**}_i $. This implies
that (\ref{eqn:lam_plus_comp}) holds in the region $[\alpha^{*}_i,-\theta_1^{-1}]$.

On the interval $\alpha \in [0, \alpha^{*}_i)$, the expression under
the square root in (\ref{eqn:roots_lam_plus}) is positive, and the
corresponding eigenvalues are real.  Thus,
\begin{align}
\max ( |\lambda_{i}^{*}(\bPhi_{3}[\alpha])|,
|\lambda_{i}^{**}(\bPhi_{3}[\alpha])| ) &=\frac{1}{2}
\left\{\begin{array}{cc} \left|\lambda_{i}(\bW_{3}[\alpha]) +
\sqrt{\lambda_{i}(\bW_{3}[\alpha])^2 + 4 \alpha \theta_{1}}\right| & \textrm{if \ } \lambda_{i}(\bW_{3}[\alpha]) \geq 0\\
\left|-\lambda_{i}(\bW_{3}[\alpha]) +
\sqrt{\lambda_{i}(\bW_{3}[\alpha])^2 + 4 \alpha \theta_{1}}\right| &
\textrm{if \ } \lambda_{i}(\bW_{3}[\alpha]) < 0,
\end{array}\right.
\end{align}
or equivalently,
%\begin{align}
$\max ( |\lambda_{i}^{*}(\bPhi_{3}[\alpha])|,
|\lambda_{i}^{**}(\bPhi_{3}[\alpha])| ) =\frac{1}{2} \left(
\left|\lambda_{i}(\bW_{3}[\alpha])\right| +
\sqrt{\lambda_{i}(\bW_{3}[\alpha])^2 + 4 \alpha \theta_{1}}
\right)$.
%\end{align}
These results establish the expression for $\mathcal{J}_i[\alpha,
\lambda_{i}(\bW)]$ in the lemma.

It remains to establish that $\mathcal{J}_1[\alpha,\lambda_1(\bW)]$
is less than all other $\mathcal{J}_i[\alpha,\lambda_i(\bW)]$ in the
region $\alpha \in [0,-\theta_1^{-1}]$.  In the region $\alpha \in
[\alpha^{*}_i, -\theta_{1}^{-1}]$, we have $-\alpha\theta_{1}^{-1}
\leq 1$, implying that $\alpha^{1/2} (-\theta_{1})^{1/2} \geq
-\alpha\theta_{1} = \mathcal{J}_1[\alpha, \lambda_{1}(\bW)]$.  In
the region $\alpha \in [0,\alpha^{*}_i)$, note that
$\lambda_{i}(\bW_{3}[\alpha])^2 + 4 \alpha \theta_{1} > 0
\Rightarrow |\lambda_{i}(\bW_{3}[\alpha])| \geq 2 (-\alpha
\theta_{1})^{1/2}$, which implies that
\begin{align}
\frac{1}{2} \left( \left|\lambda_{i}(\bW_{3}[\alpha])\right| +
\sqrt{\lambda_{i}(\bW_{3}[\alpha])^2 + 4 \alpha \theta_{1}} \right)
&\geq \frac{1}{2} \left( 2 (-\alpha \theta_{1})^{1/2} + 0 \right)
\nonumber\\
&\geq (-\alpha \theta_{1})^{1/2} \geq -\alpha \theta_{1} =\mathcal{J}_1[\alpha,
\lambda_{1}(\bW)], \label{eqn:JiversusJ1}
\end{align}
thereby establishing the final claim of the lemma.
\end{proof}

The previous lemma indicates that we can remove $\mathcal{J}_1[\alpha,
\lambda_{1}(\bW)]$ from (\ref{eqn:opt_prob_eig}), leading to a simpler optimization problem,
%\begin{align} \label{eqn:opt_prob_eig_simp}
$\alpha^{\star} = \arg \min\limits_{\alpha \geq 0}
\max\limits_{i=2,3,\ldots N} \mathcal{J}_i[\alpha, \lambda_{i}(\bW)]$.
%\end{align}
The following lemma establishes that we can simplify the optimization even further and focus solely on $\mathcal{J}_2[\alpha, \lambda_{2}(\bW)]$.

\begin{lemma} \label{lem:J2}
Under the assumptions of  Theorem~\ref{lem:optimization},
$\mathcal{J}_i[\alpha, \lambda_{i}(\bW)] \leq \mathcal{J}_2[\alpha,
\lambda_2(\bW)]$ and $\alpha^{*}_i [\lambda_{i}(\bW)] \leq
\alpha^{*}_2 [\lambda_{2}(\bW)]$ for $i=3,4,\dots,N$ over the range
$\alpha\in [0,-\theta_1^{-1}]$.
\end{lemma}

\begin{proof}
Consider the derivative of $\alpha^{*}_i [\lambda_{i}(\bW)]$ in the range
$\lambda_{i}(\bW) \in [0, 1]$:
\begin{eqnarray*}
\frac{\partial}{\partial \lambda_{i}(\bW)} \alpha^{*}_i
[\lambda_{i}(\bW)] &=&
\frac{1}{\left(\theta_{2}+\left(\theta_{3}-1\right)
\lambda_{i}(\bW)\right)^3} \times \Bigg[ \left[ 4 \theta_{1}
\left(\theta_{3}-1\right)-\theta_{2}
\left(\theta_{2}+\left(\theta_{3}-1\right) \lambda_{i}(\bW)\right) \right] \nonumber\\
& &+ \left.  \frac{\theta_{1} \left(-\theta_{2}^2+4 \theta_{1}
\left(\theta_{3}-1\right)+\theta_{2} \left(\theta_{3}-1\right)
\lambda_{i}(\bW)+2 \left(\theta_{3}-1\right)^2
\lambda_{i}(\bW)^2\right)}{\sqrt{\theta_{1}
\left(\theta_{1}+\lambda_{i}(\bW)
\left(\theta_{2}+\left(\theta_{3}-1\right)
\lambda_{i}(\bW)\right)\right)}} \right]
\end{eqnarray*}
It is clear that the multiplier outside the square brackets in the
first line above is positive in the range $\lambda_{i}(\bW) \in [0,
1]$.  Furthermore, the first summand is negative. Under the
conditions $\theta_{2} \geq 0$, $\theta_{3} \geq 1$, it can be
established that the second summand is positive and exceeds the
first summand in magnitude (see~\cite{Oresh09} for a complete
derivation). We conclude that the derivative is positive, and thus
$\alpha^{*}_i [\lambda_{i}(\bW)]$ is an increasing function over  $\lambda_{i}(\bW) \in [0, 1]$. This implies that $\alpha^{*}_i
[\lambda_{i}(\bW)] \leq \alpha^{*}_2 [\lambda_2(\bW)]$ for any $\lambda_{i} \ge 0$.

Algebraic manipulation of (\ref{eqn:alpha_roots_1}) leads to the
conclusion that $\alpha^{*}_i [-\lambda_{i}(\bW)] \leq
\alpha^{*}_i[\lambda_{i}(\bW)]$ for $\lambda_{i}(\bW) \in [0,1]$.
This implies that for positive $\lambda_{i}$, we have $\alpha^{*}_i
[-\lambda_{i}(\bW)] \leq \alpha^{*}_i [\lambda_{i}(\bW)] \leq
\alpha^{*}_2 [\lambda_2(\bW)]$. We have thus shown that
$\alpha^{*}_i [\lambda_{i}(\bW)] \leq \alpha^{*}_2 [\lambda_2(\bW)]$
for any $3 \leq i \leq N$ under the assumption $|\lambda_N(\bW)|
\leq \lambda_2(\bW)$.

\sloppypar Next we turn to proving that $\mathcal{J}_i[\alpha,
\lambda_{i}(\bW)] \leq \mathcal{J}_2[\alpha, \lambda_2(\bW)]$ for
any $3 \leq i \leq N$. We consider this problem on three distinct
intervals: $\alpha \in [0, \alpha^{*}_i[\lambda_{i}(\bW)])$, $\alpha
\in [\alpha^{*}_i[\lambda_{i}(\bW)], \alpha^{*}_2[\lambda_2(\bW)])$
and $\alpha \in [\alpha^{*}_2[\lambda_2(\bW)], -\theta_{1}^{-1}]$.
From the condition  $\alpha^{*}_i [\lambda_{i}(\bW)] \leq
\alpha^{*}_2 [\lambda_2(\bW)]$ and (\ref{eqn:partial_cost}) it is
clear that on the interval $\alpha \in
[\alpha^{*}_2[\lambda_2(\bW)], -\theta_{1}^{-1}]$ we have
$\mathcal{J}_i[\alpha, \lambda_{i}(\bW)] = \mathcal{J}_2[\alpha,
\lambda_2(\bW)] = \alpha^{1/2} (-\theta_{1})^{1/2}$. On  the
interval $\alpha \in [\alpha^{*}_i[\lambda_{i}(\bW)],
\alpha^{*}_2[\lambda_2(\bW)])$ we have $\mathcal{J}_i[\alpha,
\lambda_{i}(\bW)] = \alpha^{1/2} (-\theta_{1})^{1/2}$ and
$\mathcal{J}_2[\alpha, \lambda_2(\bW)] = \frac{1}{2} \left(
\left|\lambda_{i}(\bW_{3}[\alpha])\right| +
\sqrt{\lambda_{i}(\bW_{3}[\alpha])^2 + 4 \alpha \theta_{1}}
\right)$. From (\ref{eqn:JiversusJ1}), we see that
$\mathcal{J}_i[\alpha, \lambda_{i}(\bW)] \leq \mathcal{J}_2[\alpha,
\lambda_2(\bW)]$.

On the first interval $\alpha \in [0,
\alpha^{*}_i[\lambda_{i}(\bW)])$, we examine the
derivative of $\mathcal{J}_i[\alpha, \lambda_{i}(\bW)]$ w.r.t.
$\lambda_{i}(\bW)$:
\begin{align}
\frac{\partial}{\partial\lambda_{i}(\bW)}\mathcal{J}_i[\alpha,
\lambda_{i}(\bW)] =& \frac{1+\alpha (\theta_{3}-1)}{2}
\left(\frac{\lambda_{i}(\bW)+\alpha
\left(\theta_{2}+\left(\theta_{3}-1\right)
\lambda_{i}(\bW)\right)}{\sqrt{-4 \alpha
\left(\theta_{2}+\theta_{3}-1\right)+\left(\lambda_{i}(\bW)+\alpha
\left(\theta_{2}+\left(\theta_{3}-1\right) \lambda_{i}(\bW)\right)\right)^2}} \right. \nonumber\\
&+ \text{sgn}\left[\lambda_{i}(\bW)+\alpha
\left(\theta_{2}+\left(\theta_{3}-1\right)
\lambda_{i}(\bW)\right)\right] \Bigg)
\end{align}
We observe that the multiplier $\frac{1+\alpha (\theta_{3}-1)}{2}$
is positive, and the expression under the square root is positive
because $\alpha \in [0, \alpha^{*}_i[\lambda_{i}(\bW)])$.
Additionally, $\lambda_{i}(\bW)+\alpha
\left(\theta_{2}+\left(\theta_{3}-1\right) \lambda_{i}(\bW)\right)
\geq 0$ under the assumption $\lambda_{i}(\bW) \geq 0$ and
$\theta_{2} \geq 0$, $\theta_{3} \geq 1$. Thus
$\frac{\partial}{\partial\lambda_{i}(\bW)}\mathcal{J}_i[\alpha,
\lambda_{i}(\bW)] \geq 0$ for any $\lambda_{i}(\bW) \geq 0$ and we
have $\mathcal{J}_i[\alpha, \lambda_{i}(\bW)] \leq
\mathcal{J}_2[\alpha, \lambda_2(\bW)]$ for any $0 \leq
\lambda_{i}(\bW) \leq \lambda_2(\bW)$. Finally, we note from
(\ref{eqn:partial_cost}) that $\mathcal{J}_i[\alpha,
\lambda_{i}(\bW)]$ is an increasing function of
$|\lambda_{i}(\bW_{3}[\alpha])| =
|(1+\alpha(\theta_{3}-1))\lambda_{i}(\bW) + \alpha \theta_{2}|$.
Thus, to show that $\mathcal{J}_i[\alpha, -\lambda_{i}(\bW)] \leq
\mathcal{J}_i[\alpha, \lambda_{i}(\bW)]$ for $0 \leq
\lambda_{i}(\bW) \leq \lambda_2(\bW)$ it is sufficient to show that
$|-(1+\alpha(\theta_{3}-1))\lambda_{i}(\bW) + \alpha \theta_{2}|
\leq |(1+\alpha(\theta_{3}-1))\lambda_{i}(\bW) + \alpha
\theta_{2}|$. Under our assumptions, we have
\begin{align}
|(1+\alpha(\theta_{3}-1))\lambda_{i}(\bW) + \alpha \theta_{2}|^2 &-
|-(1+\alpha(\theta_{3}-1))\lambda_{i}(\bW) + \alpha
\theta_{2}|^2 \nonumber\\
&= 4 (1+\alpha(\theta_{3}-1))\lambda_{i}(\bW) \alpha \theta_{2} \geq
0.
\end{align}
This implies that $\mathcal{J}_i[\alpha, \lambda_{i}(\bW)] \leq \mathcal{J}_2[\alpha,
\lambda_2(\bW)]$ on the interval $\alpha \in [0,
\alpha^{*}_i[\lambda_{i}(\bW)])$, indicating that the condition
applies on the entire interval $\alpha \in [0,-\theta_1^{-1}]$, which is what we wanted to show.
\end{proof}

The remainder of the proof of Theorem~\ref{lem:optimization} proceeds as follows.
From Lemmas~\ref{lem:remJ1} and~\ref{lem:J2}, the optimization problem (\ref{eqn:opt_prob}) simplifies to:
%\begin{align} \label{eqn:opt_prob_eig_simp_fur}
$\alpha^{\star} = \arg \min\limits_{\alpha \geq 0}
\mathcal{J}_2[\alpha, \lambda_2(\bW)]$.
%\end{align}
We shall now show that $\alpha^{*}_2$ is a global minimizer of
this function. Consider the derivative of $\mathcal{J}_2[\alpha,
\lambda_2(\bW)]$ w.r.t. $\alpha$ on $[0, \alpha^{*}_2)$:
\begin{align}
\frac{\partial}{\partial\alpha}\mathcal{J}_2[\alpha, \lambda_2(\bW)]
&= \frac{2 \theta_{1}+\left(\theta_{2}+\left(\theta_{3}-1\right)
\lambda_2(\bW)\right) \left(\lambda_2(\bW)+\alpha
\left(\theta_{2}+\left(\theta_{3}-1\right)
\lambda_2(\bW)\right)\right)}{\sqrt{4 \alpha  \theta_{1}+\left(\lambda_2(\bW)+\alpha  \left(\theta_{2}+\left(\theta_{3}-1\right) \lambda_2(\bW)\right)\right)^2}} \nonumber\\
&+\left(\theta_{2}+\left(\theta_{3}-1\right) \lambda_2(\bW)\right)
\text{sgn}\left[\lambda_2(\bW)+\alpha  \left(\theta_{2}+\left(\theta_{3}-1\right) \lambda_2(\bW)\right)\right]. \nonumber
\end{align}
Denote the first term in this sum by
$\varphi_1(\lambda_2(\bW),\alpha)$ and the second by
$\varphi_2(\lambda_2(\bW),\alpha)$. It can be shown that
$|\varphi_1(\lambda_2(\bW),\alpha)|\geq
|\varphi_2(\lambda_2(\bW),\alpha)|$ for any $\lambda_2(\bW) \in [-1,
1]$ and $\alpha \in [0,\alpha^{*}_2)$ by directly solving the
inequality. We conclude that the sign of the derivative on $\alpha
\in [0, \alpha^{*}_2)$ is completely determined by the sign of
$\varphi_1(\lambda_2(\bW),\alpha)$ for $\lambda_2(\bW) \in [-1,1]$.
On $\alpha \in [0, \alpha^{*}_2)$, the sign of
$\varphi_1(\lambda_2(\bW),\alpha)$ is
 determined by the sign of its numerator. The transition point for the
numerator's sign occurs at:
$$\alpha^{+} =
-\frac{2\theta_{1}+\lambda_{2}(\bW)(\theta_{2}+(\theta_{3}-1)\lambda_{2}(\bW))}
{(\theta_{2}+(\theta_{3}-1)\lambda_{2}(\bW))^2},$$ and by showing
that $\alpha^{+} \geq -\theta_{1}^{-1}$, we can establish that this
transition point is at or beyond $\alpha_2^{*}$. This indicates that
$\varphi_1(\lambda_2(\bW),\alpha) \leq 0 \textrm{ if } \alpha \in
[0, \alpha^{*}_2)$. We observe that $\mathcal{J}_2[\alpha,
\lambda_2(\bW)]$ is nonincreasing on $\alpha \in [0, \alpha^{*}_2)$
and nondecreasing on $\alpha \in [\alpha^{*}_2, -\theta_{1}^{-1})$
(as established in Lemma~\ref{lem:remJ1}). We conclude that
$\alpha^{*}_2$ is a global minimum of the function
$\mathcal{J}_2[\alpha, \lambda_2(\bW)]$, thereby proving
Theorem~\ref{lem:optimization} and establishing
$\mathcal{J}_2[\alpha^{\star}, \lambda_2(\bW)] =
|\lambda_{2}^{*}(\bPhi_3[\alpha^{\star}])| =
\sqrt{-\alpha^{\star}\theta_{1}}$.

Note that the last argument also implies that $\mathcal{J}_2[\alpha,
\lambda_2(\bW)] \leq \lambda_2(\bW)$ on $\alpha \in [0,
\alpha^{*}_2]$ and $\mathcal{J}_2[\alpha, \lambda_2(\bW)] < 1$ on
$\alpha \in (\alpha^{*}_2, -\theta_{1}^{-1})$ since
$\mathcal{J}_2[\alpha, \lambda_2(\bW)]$ is non-increasing on the
former interval, it is non-decreasing on the latter interval and
$\mathcal{J}_2[-\theta_{1}^{-1}, \lambda_2(\bW)] = 1$. This fact
demonstrates that the matrix $\bPhi_{3}[\alpha]$ is convergent if
$\alpha \in [0, -\theta_{1}^{-1})$ in the sense that we have
$\rho(\bPhi_{3}[\alpha] - \J) < 1$.

\subsection{Proof of Theorem~\ref{th:conv_rate_bound}: Convergence Rate}
\label{ssection::conv_rate_proof}

\begin{proof}
  According to the discussion in
  Sections~\ref{ssection::OptMixResults}
  and~\ref{ssection::OptParamProof} , we have
\begin{align} \label{eqn:lambda_2_accelerated}
\rho(\bPhi_{3}[\alpha^{\star}] - \J) &=
|\lambda_{2}^*(\bPhi_3[\alpha^{\star}])| = (\alpha^{\star}
|\theta_{1}|)^{1/2} \nonumber\\
&= \left[ \frac{-((\theta_{3}-1) \lambda_2^2 + \theta_{2} \lambda_2
+ 2\theta_{1}) - 2\sqrt{\theta_{1}^2 +
\theta_{1}\lambda_2\left(\theta_{2}+(\theta_{3}-1)\lambda_2
\right)}}{\left(\theta_{2} + (\theta_{3}-1)\lambda_2\right)^2 }
|\theta_{1}| \right]^{1/2}. \nonumber
\end{align}
In order to prove the claim, we consider two cases: $\lambda_2(\bW) = 1 - \Psi(N)$, and $\lambda_2(\bW) < 1 - \Psi(N)$.

First, we suppose that $\lambda_{2}(\bW) = 1 - \Psi(N)$ and show
that $\rho(\bPhi_3[\alpha^{\star}] - \J)^2 - (1 - \sqrt{\Psi(N)})^2
\leq 0$. Denoting $\Psi(N) = \delta$ and substituting
$\lambda_{2}(\bW) = 1 - \delta$ and $\theta_1 = 1 - \theta_2 -
\theta_3$, we obtain
\begin{align*}
\rho(\bPhi_3[\alpha^{\star}] - \J)^2 &- (1 - \sqrt{\Psi(N)})^2 =
-\left(\sqrt{\delta }-1\right)^2 \nonumber \\
&\times \frac{\left(\theta_{3}-1\right)(\delta ^2-\delta) + 2
\sqrt{\delta } \left(\theta_{3}+\theta_{2}-1\right)-2 \sqrt{\delta
\left(\theta_{2}+(2-\delta) \left(\theta_{3}-1\right)\right)
\left(\theta_{3}+\theta_{2}-1\right)}}{[(2-\delta )\delta +
1](1-\theta_{3}) -(1+\delta ) \theta_{2}-2 \sqrt{\delta
\left(\theta_{3}+\theta_{2}-1\right)
\left((\theta_{3}-1)(2-\delta)+\theta_{2}\right)}}.
\end{align*}
It is clear from the assumptions that the expressions under square roots are
non-negative.  Furthermore, the denominator is negative since
$1-\theta_{3} < 0$, $\theta_{2} > 0$ and $\delta \in (0,1)$.
Finally, note that $\left(\theta_{3}-1\right)(\delta ^2-\delta) \leq
0$ and $2 \sqrt{\delta } \left(\theta_{3}+\theta_{2}-1\right) \geq
0$. Thus, to see that the numerator is non-positive, observe that
\begin{align}
[\sqrt{\delta } \left(\theta_{3}+\theta_{2}-1\right)]^2 &- \left[
\sqrt{\delta \left(\theta_{2}+(2-\delta)
\left(\theta_{3}-1\right)\right)
\left(\theta_{3}+\theta_{2}-1\right)} \right]^2 \nonumber\\
&= (\delta - 1)\delta(\theta_{3}-1)(\theta_{3}+\theta_{2}-1) \leq 0.
\end{align}
Thus, we have $\rho(\bPhi_3[\alpha^{\star}] - \J)^2 - (1 -
\sqrt{\Psi(N)})^2 \leq 0$, implying that
$\rho(\bPhi_3[\alpha^{\star}] - \J) \leq 1 - \sqrt{\Psi(N)}$ if
$\lambda_{2}(\bW) = 1 - \Psi(N)$.

Now suppose $\lambda_{2}(\bW) < 1 - \Psi(N)$. We have seen in
Lemma~\ref{lem:J2} that $\alpha_{i}^{*}[\lambda_i(\bW)]$ is an
increasing function of $\lambda_{i}(\bW)$, implying
$\alpha_{2}^{*}[\lambda_{2}(\bW)] \leq \alpha_{2}^{*}[1 - \Psi(N)]$.
Since $\rho(\bPhi_3[\alpha^{\star}] - \J) = (\alpha^{\star}
|\theta_{1}|)^{1/2} =
(\alpha_{2}^{*}[\lambda_{2}(\bW)]|\theta_{1}|)^{1/2}$ is an
increasing function of $\alpha_{2}^{*}[\lambda_{2}(\bW)]$,  the claim of theorem follows.
\end{proof}

\subsection{Proof of Theorem~\ref{th:t_lim_ratio}: Expected Gain}
\label{ssection::proof_processing_gain}

\begin{proof}
First, condition on a particular realization of the graph topology, and observe from the definition of $\tau_{\asym}(\cdot)$ that
\begin{align}
\frac{\tau_{\asym}(\bW)}{\tau_{\asym}(\bPhi_3[\alpha^{\star}])} =
\frac{\log \rho(\bPhi_3[\alpha^{\star}] - \J)}{\log \rho(\bW - \J)}.
\end{align}
Next, fixing $\rho(\bW - \J) = 1 - \psi$, where $\Psi(N) =
\E\{\psi\}$, and using Theorem~\ref{th:conv_rate_bound}, we have
\begin{align}
\frac{\tau_{\asym}(\bW)}{\tau_{\asym}(\bPhi_3[\alpha^{\star}])} \geq
\frac{\log( 1 - \sqrt{\psi})}{\log( 1 - \psi)}.
\end{align}
Let $f(x) = \log( 1 - \sqrt{x})/\log( 1 - x)$. Taking the Taylor
series expansion of $f(x)$ at $x = 0$, we obtain
\begin{align}
f(x) = \frac{1}{\sqrt{x}} + \frac{1}{2} - \frac{1}{6} x^{1/2} -
\frac{1}{20} x^{3/2} - \ldots.
\end{align}
Noting that $x > 0$ we conclude that the following holds uniformly
over $x \in [0, 1]$:
%\begin{align}
$f(x) \leq \frac{1}{\sqrt{x}} + \frac{1}{2}$.
%\end{align}
At the same time, taking the Taylor series expansions of the numerator and
denominator of $f(x)$, we obtain
\begin{align}
f(x) = \frac{\sqrt{x} + \frac{x}{2} + \frac{x^{3/2}}{3}  +
\frac{x^{2}}{4}  + \frac{x^{5/2}}{5} + \ldots}{ x + \frac{x^2}{2} +
\frac{x^{3}}{3}  + \ldots }.
\end{align}
Noting that $1/6+1/3=1/2$, $2/15+1/5=1/3$, we can express this as
\begin{align}
f(x) = \frac{1}{\sqrt{x}} \frac{\sqrt{x} + \frac{x}{2} +
\frac{x^{3/2}}{3} + \frac{x^{2}}{4}  + \frac{x^{5/2}}{5} + \ldots}{
\sqrt{x} + \frac{x^{3/2}}{6} + \frac{x^{3/2}}{3} +
\frac{2x^{5/2}}{15} + \frac{x^{5/2}}{5}  + \ldots },
\end{align}
and using the fact that $1/2x \geq 1/6 x^{3/2}$, $1/4 x^2 \geq 2/15
x^{5/2}$ , $\ldots$ uniformly over $x \in [0, 1]$, we conclude that
$f(x) \geq \frac{1}{\sqrt{x}}$.  Thus, $\frac{1}{\sqrt{x}} \leq f(x)
\leq \frac{1}{\sqrt{x}} + \frac{1}{2}$, where both bounds are tight.
Finally, observe that $\frac{\partial^2}{\partial x^2} x^{-1/2} =
3/4 x^{-5/2}
> 0$ if $ x > 0$, implying that $1/\sqrt{x}$ is convex.  To complete the proof we take
the expectation with respect to graph realizations and apply Jensen's
inequality to obtain
\begin{align}
\E\left\{
\frac{\tau_{\asym}(\bW)}{\tau_{\asym}(\bPhi_3[\alpha^{\star}])}
\right\} \geq \E\left\{ \frac{1}{\sqrt{\psi}} \right\} \geq
\frac{1}{\sqrt{\E\left\{\psi\right\}}}.
\end{align}
\end{proof}

\section{Concluding Remarks}
\label{section::Concluding Remarks}

This paper provides theoretical performance guarantees for
accelerated distributed averaging algorithms using node memory.  We
consider acceleration based on local linear prediction and focus on
the setting where each node uses two memory taps.  We derived the
optimal value of the mixing parameter for the accelerated averaging
algorithm and discuss a fully-decentralized scheme for estimating
the spectral radius, which is then used to initialize the optimal
mixing parameter. An important contribution of this paper is the
derivation of upper bounds on the spectral radius of the accelerated
consensus matrix.  This bound relates the spectral radius growth
rate of the original matrix with that of the accelerated consensus
matrix. We believe that this result applies to the general class of
distributed averaging algorithms using node state prediction, and
shows that, even in its simplified form and even at the theoretical
level, accelerated consensus may provide considerable processing
gain. We conclude that this gain, measured as the ratio of the
asymptotic averaging time of the non-accelerated and accelerated
algorithms, grows with increasing network size. Numerical
experiments confirm our theoretical conclusions and reveal the
feasibility of online implementation of the accelerated algorithm
with nearly optimal properties. Finding ways to analyze the proposed
algorithm in more general instantiations and proposing simpler
initialization schemes are the focus of ongoing investigation.

\bibliographystyle{IEEEtran}
\bibliography{IEEEabrv,additional}

\textbf{Copyright (c) 2010 IEEE. Personal use of this material is
permitted. However, permission to use this material for any other
purposes must be obtained from the IEEE by sending a request to
pubs-permissions@ieee.org.}

\end{document}